\documentclass[9.5pt,compsoc, cspaper]{IEEEtran}
\usepackage{array}
\usepackage{stfloats}
\usepackage{verbatim}
\usepackage{cite}
\usepackage{amsmath,amssymb,amsfonts}
\usepackage{algorithmic}
\usepackage{graphicx}
\usepackage{textcomp}
\usepackage{multirow}
\usepackage{booktabs}
\usepackage[inline]{enumitem}
\usepackage[table, xcdraw]{xcolor}
\usepackage{subcaption}
\usepackage{scalerel}
\usepackage{tikz}
\usetikzlibrary{svg.path}
\definecolor{orcidlogocol}{HTML}{A6CE39}
\tikzset{
  orcidlogo/.pic={
    \fill[orcidlogocol] svg{M256,128c0,70.7-57.3,128-128,128C57.3,256,0,198.7,0,128C0,57.3,57.3,0,128,0C198.7,0,256,57.3,256,128z};
    \fill[white] svg{M86.3,186.2H70.9V79.1h15.4v48.4V186.2z}
                 svg{M108.9,79.1h41.6c39.6,0,57,28.3,57,53.6c0,27.5-21.5,53.6-56.8,53.6h-41.8V79.1z M124.3,172.4h24.5c34.9,0,42.9-26.5,42.9-39.7c0-21.5-13.7-39.7-43.7-39.7h-23.7V172.4z}
                svg{M88.7,56.8c0,5.5-4.5,10.1-10.1,10.1c-5.6,0-10.1-4.6-10.1-10.1c0-5.6,4.5-10.1,10.1-10.1C84.2,46.7,88.7,51.3,88.7,56.8z};
 }
}
\newcommand\orcidicon[1]{\href{https://orcid.org/#1}{\mbox{\scalerel*{
\begin{tikzpicture}[yscale=-1,transform shape]
\pic{orcidlogo};
\end{tikzpicture}
}{|}}}}

\usepackage{hyperref} 
\usepackage{cleveref}

\newcommand{\round}[1]{\ensuremath{\lfloor#1\rceil}}
\def\BibTeX{{\rm B\kern-.05em{\sc i\kern-.025em b}\kern-.08em
    T\kern-.1667em\lower.7ex\hbox{E}\kern-.125emX}}
\usepackage{balance}

\begin{document}
\title{FrankenSplit: Efficient Neural Feature Compression with Shallow Variational Bottleneck Injection for Mobile Edge Computing}
\author{Alireza Furtuanpey \orcidicon{0000-0001-5621-7899}, Philipp Raith \orcidicon{0000-0003-3293-9437}, Schahram Dustdar \orcidicon{ 0000-0001-6872-8821}, \textit{Fellow, IEEE}}


\IEEEtitleabstractindextext{%
 \begin{abstract}
The rise of mobile AI accelerators allows latency-sensitive applications to execute lightweight Deep Neural Networks (DNNs) on the client side. However, critical applications require powerful models that edge devices cannot host and must therefore offload requests, where the high-dimensional data will compete for limited bandwidth.
Split Computing (SC) alleviates resource inefficiency by partitioning DNN layers across devices, but current methods are overly specific and only marginally reduce bandwidth consumption.
This work proposes shifting away from focusing on executing shallow layers of partitioned DNNs.
Instead, it advocates concentrating the local resources on variational compression optimized for machine interpretability.
We introduce a novel framework for resource-conscious compression models and extensively evaluate our method in an environment reflecting the asymmetric resource distribution between edge devices and servers.
Our method achieves 60\% lower bitrate than a state-of-the-art SC method without decreasing accuracy and is up to 16x faster than offloading with existing codec standards.
\end{abstract}
}
\maketitle
\IEEEdisplaynontitleabstractindextext

\section{Introduction} \label{sec:intro}
\IEEEPARstart{D}{eep} Learning (DL) has demonstrated that it can solve real-world problems in challenging areas ranging from Computer Vision (CV)~\cite{voulodimos2018deep} to Natural language Processing (NLP)~\cite{otter2020survey}. Complementary with the advancements in mobile edge computing (MEC)~\cite{feng2022computation} and energy-efficient AI accelerators, visions of intelligent city-scale platforms for critical applications, such as mobile augmented reality (MAR)~\cite{rauschcogxr}, disaster warning~\cite{disasterwarning}, or facilities management~\cite{xin2022deep}, seem progressively feasible. Nevertheless, the accelerating pervasiveness of mobile clients gave unprecedented growth in Machine-to-Machine (M2M) communication~\cite{cisco2020cisco}, leading to an insurmountable amount of network traffic. A root cause is the intrinsic limitation of mobile devices that allows them to realistically host a single lightweight Deep Neural Network (DNN) in memory at a time. Local resources cannot meet the demanding requirements of applications that rely on multiple highly accurate DNNs~\cite{zhang2022comprehensive, zhang2023comprehensive}. Hence, clients must frequently offload inference requests~~\cite{infaas}.

The downside to offloading is that by constantly streaming high-dimensional visual data, the limited bandwidth will inevitably lead to network congestion, resulting in erratic response delays, and it leaves valuable client-side resources idle.

Split Computing (SC) emerged as an alternative to alleviate inefficient resource utilization and to facilitate low-latency and performance-critical mobile inference. The basic idea is to partition a DNN to process the shallow layers with the client and send a processed representation to the remaining deeper layers deployed on a server. The SC paradigm can potentially draw resources from the entire edge-cloud compute continuum. However, current SC methods are only conditionally applicable (e.g., in highly bandwidth-constrained networks) or tailored toward specific neural network architectures. Methods that claim to generalize towards a broader range of architectures do not consider that mobile clients can typically only load a single model into memory. Consequently, SC methods are impractical for applications with complex requirements relying on inference from multiple models concurrently (e.g., MAR).  Mobile clients reloading weights from its storage into memory and sending multiple intermediate representations for each pruned model would incur more overhead than directly transmitting image data with fast lossless codecs. Moreover, due to the conditional applicability of SC, practical methods rely on a decision mechanism that periodically probes external conditions (e.g., available bandwidth), resulting in further deployment and runtime complexity~\cite{matsubara2022split}. 

This work shows that we can address the increasing need to reduce bandwidth consumption while simultaneously generalizing the objective of SC methods to provide mobile clients access to low-latency inference from remote off-the-shelf discriminative models even in constrained networks.

We draw from recent advancements in lossy learned image compression (LIC) and the Information Bottleneck (IB) principle~\cite{informationbottleneck}. Despite outperforming handcrafted codecs~\cite{introneuralcompression}, such as PNG, or WebP~\cite{webp}, LIC is unsuitable for real-time inference in MEC since they consist of large models and other complex mechanisms that are demanding even for server-grade hardware. Further, research in compression primarily focuses on reconstruction for human perception containing information superfluous for M2M communication. In comparison, the deep variational information bottleneck (DVIB) provides an objective for learned feature compression with DNNs, prioritizing information valuable for machine interpretability.

With DVIB, we can conceive generalizable methods that are applicable to off-the-shelf architectures. However, current DVIB approaches typically place the bottleneck at the penultimate layer. Thus, they are unsuitable for most common settings that assume an asymmetric resource allocation between the client and the server. In other words, the objectives of DVIB and MEC contradict each other, i.e., for the latter, we require shifting the bottleneck's location to the shallow layers.

We accommodate the restrictions of mobile clients by introducing a method that moves the bottleneck to the shallow layers and retains generalizability to arbitrary architectures. While shifting the bottleneck does not formally change the objective, we will demonstrate that existing methods for mutual information estimation lead to unsatisfactory results. 

To this end, we introduce \textit{FrankenSplit}: A novel training and design heuristic for variational feature compression models embeddable in arbitrary DNN architectures with pre-trained weights for high-level vision tasks. FrankenSplit is refreshingly simple to implement and deploy without additional decision mechanisms that rely on runtime components for probing external conditions. Additionally, by deploying a single lightweight encoder, the client can access state-of-the-art accuracy from multiple large server-grade models without reloading weights from memory for each task. Lastly, the approach does not require modifying discriminative models (e.g., by finetuning weights). Therefore, we can directly utilize foundational off-the-shelf models and seamlessly integrate FrankenSplit into existing systems.

We open-source our repository \footnote{https://github.com/rezafuru/FrankenSplit} as an addition to the community for researchers to reproduce and extend our experiments. In summary, our contributions are:
\begin{itemize}
\item Thoroughly exploring how shallow and deep bottleneck injection differ for feature compression. 
 \item Introducing a novel saliency-guided training method to overcome the challenges of training a lightweight encoder with limited capacity to compress features usable for several downstream tasks. 
\item Introducing a generalizable design heuristic for embedding a variational feature compression model into arbitrary DNN architectures.
\end{itemize}
\Cref{sec:relwork} discusses relevant work on SC and LIC. \Cref{sec:casefornfc} discusses the limitations of SC methods and motivates neural feature compression.
\Cref{sec:probform} describes the problem domain. \Cref{sec:solapproach} progressively introduces the solution approach. \Cref{sec:eval} extensively justifies relevant performance indicators and evaluates several implementations of FrankenSplit against various baselines to assess our method's efficacy. Lastly, \Cref{sec:conc} summarizes this work and highlights limitations to motivate follow-up work.
\section{Related Work} \label{sec:relwork}
\subsection{Neural Data Compression} \label{subsec:neuralcompression}
\subsubsection{Learned Image Compression} \label{subsubsec:imagecompression}
The goal of (lossy) image compression is minimizing bitrates while preserving information critical for human perception. Transform coding is a basic framework of lossy compression, which divides the compression task into decorrelation and quantization~\cite{transformcoding2001}. Decorrelation reduces the statistical dependencies of the pixels, allowing for more effective entropy coding, while quantization represents the values as a finite set of integers. The core difference between handcrafted and learned methods is that the former relies on linear transformations based on expert knowledge. Contrarily, the latter is data-driven with non-linear transformations learned by neural networks~\cite{nonlineartc2020}. 

Ballé et al. introduced the Factorized Prior (FP) entropy model and formulated the neural compression problem by finding a representation with minimal entropy~\cite{balle2017}. An encoder network transforms the original input to a latent variable, capturing the input’s statistical dependencies. In follow-up work, Ballé et al.~\cite{balle2018} and Minnen et al.~\cite{minnen2018} extend the FP entropy model by including a hyperprior as side information for the prior. Minnen et al.~\cite{minnen2018} introduce the joint hierarchical priors and autoregressive entropy model (JHAP), which adds a context model to the existing scale hyperprior latent variable models. Typically, context models are lightweight, i.e., they add a negligible number of parameters, but their sequential processing increases the end-to-end latency by orders of magnitude. 
\subsubsection{Feature Compression} \label{subsubsec:featurecompression}
Singh et al. demonstrate a practical method for the Information Bottleneck principle in a compression framework by introducing the bottleneck in the penultimate layer and replacing the distortion loss with the cross-entropy for image classification~\cite{singh2020}. Dubois et al. generalized the VIB for multiple downstream tasks and were the first to describe the feature compression task formally~\cite{dubois2021}. However, their encoder-only CLIP compressor has over 87 million parameters. Both Dubois and Singh et al. consider feature compression for mass storage, i.e., they assume the data is already present at the target server. In contrast, we consider how resource-constrained clients must first compress the high-dimensional visual data before sending it over a network. 

Closest to our work is the Entropic Student (ES) proposed by Matsubara et al.~\cite{sc2bench, matsubara2022supervised}, as we follow the same objective of real-time inference with feature compression. Nevertheless, they simply apply the learning objective and a scaled-down version of autoencoder from \cite{balle2017, balle2018}. 
Contrastingly, we carefully examine the problem domain of resource-conscious feature compression to identify underlying issues with current methods, allowing us to conceive novel solutions with significantly better rate-distortion performance.
\subsection{Split Computing}\label{subsec:splitcomputing}
 We distinguish between two orthogonal approaches to SC.
\subsubsection{Split Runtimes}\label{subsubsec:splitruntimes} Split runtime systems are characterized by performing no or minimal modifications on off-the-shelf DNNs. The objective is to dynamically determine split points according to the available resources, network conditions, and intrinsic model properties. Hence, split runtimes primarily focus on profilers and adaptive schedulers. Kang et al. performed extensive compute cost and feature size analysis on the layer-level characterizations of DNNs and introduced the first split runtime system~\cite{neurosurgeon2017}. Their study has shown that split runtimes are only sensible for DNNs with an early natural bottleneck, i.e., models performing aggressive dimensionality reduction within the shallow layers. However, most modern DNNs increase feature dimensions until the last layers for better representation. Consequently, follow-up work focuses on feature tensor manipulation~\cite{jalad2018, spinn2020, dyno2022}. We argue against split runtimes since they introduce considerable complexity. Worse, the system must be tuned toward external conditions, with extensive profiling and careful calibration. Additionally, runtimes raise overhead and another point of failure by hosting a network-spanning system. Notably, even the most sophisticated methods still rely on a natural bottleneck, evidenced by how state-of-the-art split runtimes still report results on superseded DNNs with an early bottleneck~\cite{loadpart2022, bakhtiarnia2023}.
\subsubsection{Artificial Bottleneck Injection}\label{subsubsec:abi} By shifting the effort towards modifying and re-training an existing base model (backbone) to replace the shallow layers with an artificial bottleneck, bottleneck injection retains the simplicity of offloading. Eshratifar et al. replace the shallow layers of ResNet-50 with a deterministic autoencoder network~\cite{bottlenet2019}. A follow-up work by Jiawei Shao and Jun Zhang further considers noisy communication channels~\cite{bottlenetpp2020}. Matsubara et al.~\cite{matsubarahnd2019}, and Sbai et al.~\cite{cde2021} propose a more general network agnostic knowledge distillation (KD) method for embedding autoencoders, where the output of the split point from the unmodified backbone serves as a teacher. Lastly, we consider the work in~\cite{sc2bench} as the state-of-the-art for bottleneck injection.

Although bottleneck injection is promising, there are two problems with current methods. They rely on deterministic autoencoders for a crude approximation to compression or are intended for a specific class of neural network architecture.

This work addresses both limitations of such bottleneck injection methods. 
\section{The Case for Neural Data Compression}\label{sec:casefornfc}
We assume an asymmetric resource allocation between the client and the server, i.e., the latter has considerably higher computational capacity. Additionally, we consider large models for state-of-the-art performance of non-trivial discriminative tasks unsuitable for mobile clients. 
Progress in energy-efficient ASICs and embedded AI with model compression with quantization, channel pruning, etc., permit constrained clients to execute lightweight DNNs. Nevertheless, they are bound to reduced predictive strength relative to their contemporary unconstrained counterparts~\cite{modelcompressionsurvey}. This assumption is sensible considering the trend for DNNs towards pre-trained foundational models with rising computational requirements due to increasing model sizes~\cite{cnnsurvey} and costly operations~\cite{transformersurvey}. 

Lastly, mobile devices cannot realistically load weights for multiple models simultaneously~\cite{zhang2023comprehensive}, and it is unreasonable to expect that a single compressed model is sufficient for applications with complex requirements that rely on various models concurrently or in quick succession. 

Conclusively, despite the wide availability of onboard accelerators, the demand for large models to solve intelligent tasks will further increase, transmitting large volumes of high-dimensional data. The claim is consistent with CISCO's report that emphasizes the accelerating bandwidth consumption by M2M communication~\cite{cisco2020cisco}.
\subsection{Limitations of Split Computing} \label{subsec:limsc}
\begin{figure}[htbp]
    \centering
    \includegraphics[width=\columnwidth]{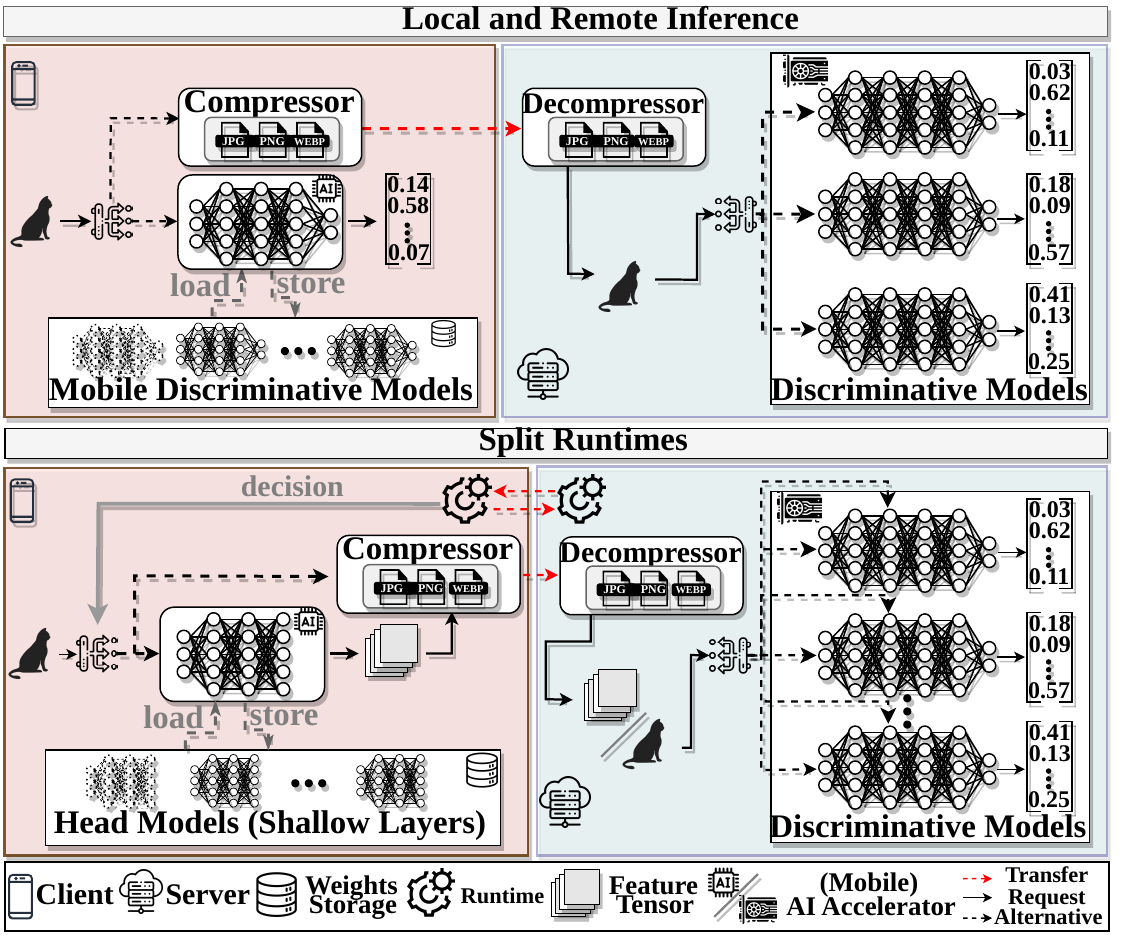}
    \caption{Prediction with on/offloading and split runtimes}
    \label{fig:infstrategiesppt1}
\end{figure}

Still, it would be valuable to leverage advancements in energy-efficient mobile chips beyond applications where local inference is sufficient. In particular, SC can potentially draw resources from an entire edge-cloud compute continuum while binary on- or offloading decision mechanisms will leave valuable client or server-side resources idle. \Cref{fig:infstrategiesppt1} illustrates generic on/offloading and split runtimes. 
The caveat is that both SC approaches discussed in \Cref{subsec:splitcomputing} are only conditionally applicable.  In particular, split runtimes reduce server-side computation for inference tasks with off-the-shelf models by onloading and executing shallow layers at the client. This approach introduces two major limitations.

First, when the latency is crucial, this is only sensible if the time for client-slide execution, transferring the features, and remotely executing the remaining layers is less than the time of directly offloading the task.  More recent work~\cite{loadpart2022, dyno2022, bakhtiarnia2023} relies on carefully calibrated dynamic decision mechanisms. A runtime component periodically measures (e.g., network bandwidth) and internal conditions (e.g., client load) to measure ideal split points or whether direct offloading is preferable. 

Second, since the shallow layers must match the deeper layers, split runtimes cannot accommodate applications with complex requirements, which is a common justification for MEC (e.g., MAR). Constrained clients would need to swap weights from the storage in memory each time the prediction model changes. Worse, the layers must match even for models predicting the same classes with closely related architectures. 

Hence, it is particularly challenging to integrate split runtimes into systems that can increase the resource efficiency of servers by adapting to shifting and fluctuating environments~\cite{romero2021infaas, reason2}. For example, when a client specifies a target accuracy and a tolerable lower bound, the system could select a ResNet-101 that can hit the target accuracy but may temporarily fall back to a ResNet-50 to ease the load when necessary. 
\subsection{Execution Times with Resource Asymmetry} \label{subsec:execasym}
\Cref{tab:sccompdist} summarizes the results of a simple experiment to demonstrate limitations incurred by resource asymmetry. The client is an Nvidia Jetson NX2 equipped with an AI accelerator, and the server hosts an RTX 3090 (see \Cref{sec:eval} for details on hardware configurations). We measure the execution times of ResNet variants, classifying a single $3\times 224 \times 224$ tensor at two split points. \begin{table}[htb]
\centering
\caption{Execution Times of Split Models}
\label{tab:sccompdist}
\resizebox{\columnwidth}{!}{%
\begin{tabular}{ccccccc}
\hline
Model &
  \begin{tabular}[c]{@{}c@{}}Split\\ Index\end{tabular} &
  \begin{tabular}[c]{@{}c@{}}Head \\ {[}NX2{]} (ms)\end{tabular} &
  \begin{tabular}[c]{@{}c@{}}Head \\ {[}3090{]} (ms)\end{tabular} &
  \begin{tabular}[c]{@{}c@{}}Tail \\ {[}3090{]} (ms)\end{tabular} &
  \begin{tabular}[c]{@{}c@{}}Rel. Exec.\\  {[}NX2{]} (\%)\end{tabular} &
  \begin{tabular}[c]{@{}c@{}}Contribution \\ {[}NX2{]} (\%)\end{tabular} \\ \hline
 &
  Stem &
  1.5055 &
  0.1024 &
  4.9687 &
  23.25 &
  0.037 \\
\multirow{-2}{*}{ResNet-50} &
  \cellcolor[HTML]{EDEDED}Stage 1 &
  \cellcolor[HTML]{EDEDED}8.2628 &
  \cellcolor[HTML]{EDEDED}0.9074 &
  \cellcolor[HTML]{EDEDED}4.0224 &
  \cellcolor[HTML]{EDEDED}67.26 &
  \cellcolor[HTML]{EDEDED}0.882 \\ \hline
 &
  Stem &
  1.5055 &
  0.1024 &
  9.8735 &
  13.23 &
  0.021 \\
\multirow{-2}{*}{ResNet-101} &
  \cellcolor[HTML]{EDEDED}Stage 1 &
  \cellcolor[HTML]{EDEDED}8.2628 &
  \cellcolor[HTML]{EDEDED}0.9074 &
  \cellcolor[HTML]{EDEDED}8.9846 &
  \cellcolor[HTML]{EDEDED}47.91 &
  \cellcolor[HTML]{EDEDED}0.506 \\ \hline
 &
  Stem &
  1.5055 &
  0.1024 &
  14.8862 &
  9.18 &
  0.015 \\
\multirow{-2}{*}{ResNet-152} &
  \cellcolor[HTML]{EDEDED}Stage 1 &
  \cellcolor[HTML]{EDEDED}8.2628 &
  \cellcolor[HTML]{EDEDED}0.9074 &
  \cellcolor[HTML]{EDEDED}13.8687 &
  \cellcolor[HTML]{EDEDED}37.34 &
  \cellcolor[HTML]{EDEDED}0.374 \\ \hline
\end{tabular}%
}
\end{table}

Similar to other widespread architectural families, ResNets organize their layers into four top-level layers, and the top-grained ones recursively consist of finer-grained ones. While the terminology differs for architectures, we will uniformly refer to top-level layers as \textit{stages} and the coarse-grained layers as \textit{blocks}.

Split point \textit{stem} assigns the first preliminary block as the head model. It consists of a convolutional layer with batch normalization~\cite{ioffe2015batch} and ReLU activation, followed by max pooling. Split point \textit{Stage 1} additionally assigns the first stage to the head. 
Notice how the shallow layers barely constitute the overall computation, even when the client takes more time to execute the head than the server for the entire model. Further, compare the percentage of total computation time and relate them to the number of parameters. At best, the client contributes to 0.02\%  of the model execution when taking 9\% of the total computation time and may only contribute 0.9\% when taking 67\% off the computation time. 

Despite a powerful AI accelerator, it is evident that utilizing client-side resources to aid a server is inefficient. Consequently, SC methods commonly include some form of quantization and data size reduction to offset resource asymmetry. In the following, we conceive a hypothetical SC method to provide intuition behind the importance of reducing transfer costs. 
\subsection{Feature Tensor Dimensionality and Quantization} \label{subsec:ftdimquant}
\begin{figure}[htbp]
    \centering
    \includegraphics[width=\columnwidth]{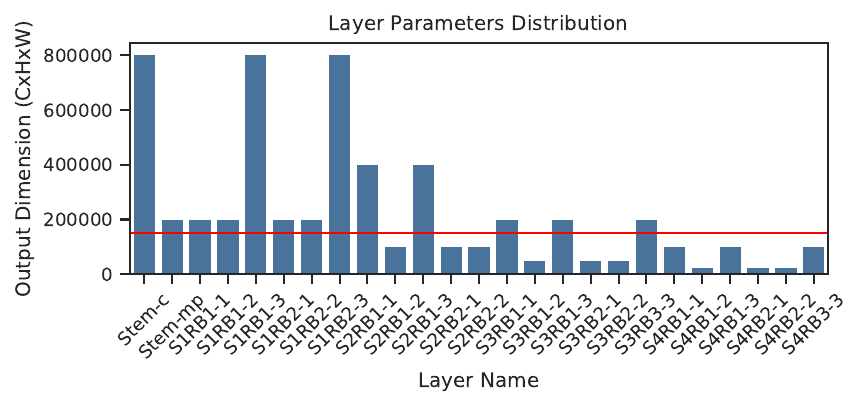}
    \caption{Output dimensionality distribution for ResNet}
    \label{fig:layer_distribution}
\end{figure}

\begin{figure}[htbp]
    \centering
    \includegraphics[width=\columnwidth]{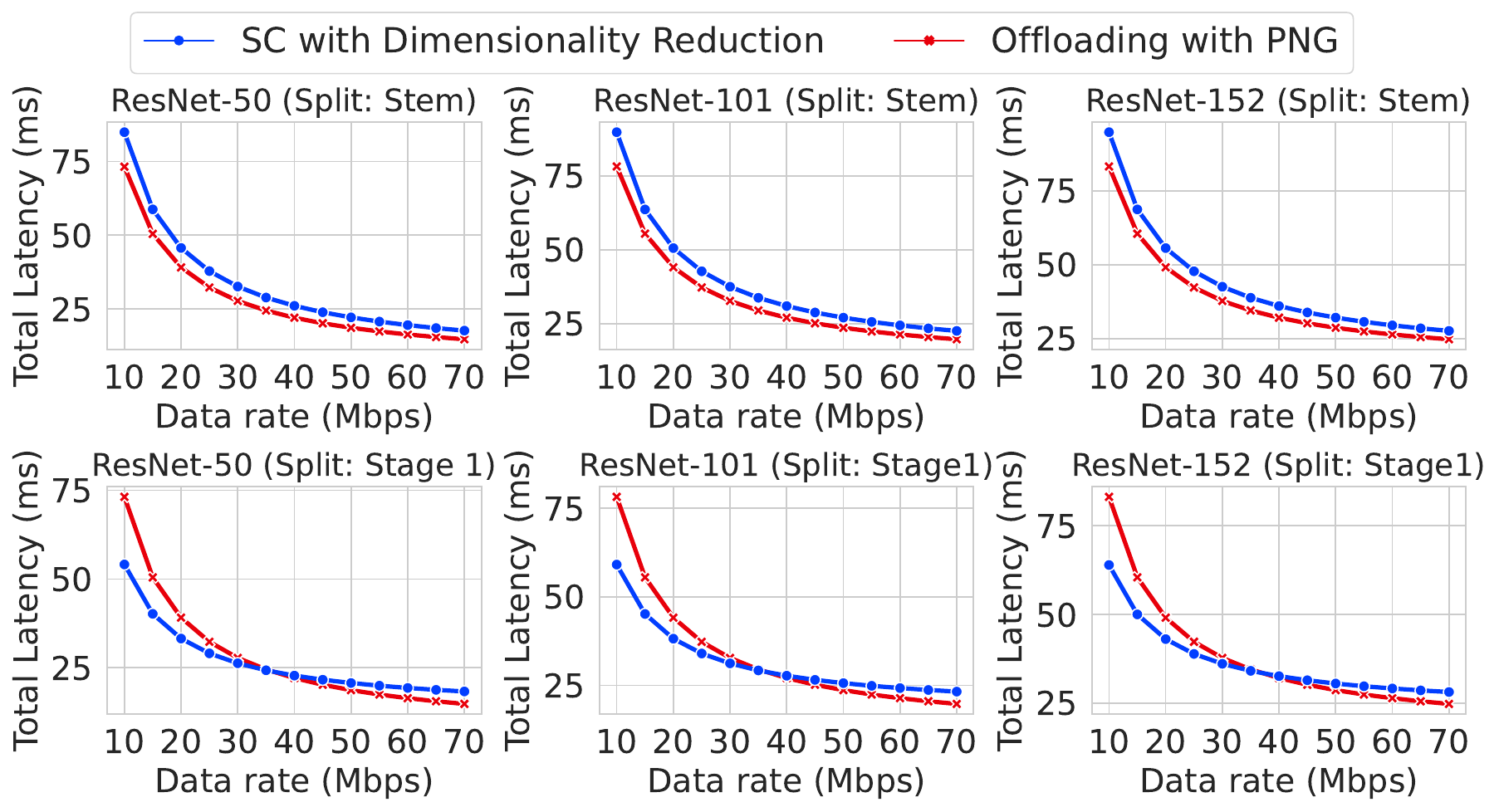}
    \caption{Inference latency for SC and offloading}
    \label{fig:splitvsoffload}
\end{figure}

Typically, most work starts with some statistical analysis of the output layer dimensions, as illustrated in \Cref{fig:layer_distribution}. Excluding repeating blocks, the feature dimensionality is identical for ResNet-50, -101, and -152. The red line marks the cutoff where the size of the intermediate feature tensor is less than the original input. ResNets (including more modern variants~\cite{resnext}), among numerous recent architectures~\cite{cnnsurvey, transformersurvey}, do not have an early natural bottleneck and will only drop below the cutoff from the first block of the second stage (S3RB1-2). Since executing until S3RB1-2 is only about 0.06\% 
Modern methods reduce the number of layers a client must execute with feature tensor quantization and other clever (typically statistical) methods that statically or dynamically prune channels~\cite{matsubara2022split}. 
For our hypothetical method, we use the execution times from \Cref{tab:sccompdist}. We generously assume that the method applies feature tensor quantization and channel pruning to reduce the expected data size without a loss in accuracy for the \textit{ImageNet} classification task~\cite{imagenet} and with no computational costs. Further, we reward the client for executing deeper layers to reflect deterministic bottleneck injection methods, such that the output size of the stem and stage one are $802816$ and $428168$ bits, respectively. Note that, for stage one, this is roughly a 92\% reduction relative to its original FP32 output size. Yet, the plots in \Cref{fig:splitvsoffload} show that offloading with PNG, let alone more modern lossless codecs (e.g., WebP), will beat SC in total request time, except when the data rate is severely constrained.
Evidently, using reasonably powerful AI accelerators to execute the shallow layers of a target model is not an efficient use of client-side resources.
\subsection{The advantage of learned methods} \label{subsec:advlearned}
In a narrow sense, more modern work on SC considers minimizing transmitting data with feature tensor quantization and other clever (typically statistical) methods that statically or dynamically prune channels.

While dimensionality reduction can be seen as a crude approximation to compression, it is not equivalent to it~\cite{minnen2018}. Compression aims to reduce the entropy of the latent under a prior shared between the sender and the receiver~\cite{nonlineartc2020}. Dimensionality reduction (especially channel pruning) may seem effective for simple tasks (e.g., CIFAR-10~\cite{krizhevsky2009learning}). However, this is more due to the overparameterization of large DNNs. Precisely, for a simple task, we can prune most channels or inject a small autoencoder at the shallow layers that may appear to achieve unprecedented compression rates relative to the unmodified head’s feature tensor size. In \Cref{subsubsec:naivetydo}, we will show that methods working reasonably well on a simple dataset can ultimately falter on more challenging datasets.

From an information-theoretic point of view~\cite{shwartz2017opening}, tensor dimensionality is not an adequate measure (i.e., $C \times H \times W \times \text{Precision}$) to determine transfer costs. Instead, we should consider the entropy of the feature tensor~\cite{shwartz2017opening}. 
Then, we can optimize a model to reduce uncertainty and compress an input according to its information content.

To summarize, the potential of SC is inhibited by primarily focusing on shifting parts of the model execution from the server to the client. SC's viability is not determined by how well they can partially compute a split network but by how well they can reduce the input size. Therefore, we pose the following question: \textit{Is it more efficient to focus the local resources \textbf{exclusively} on compressing the data rather than executing shallow layers of a network that would constitute a negligible amount of the total computation cost on the server?}

\begin{figure}[htbp]
    \centering
    \includegraphics[width=\columnwidth]{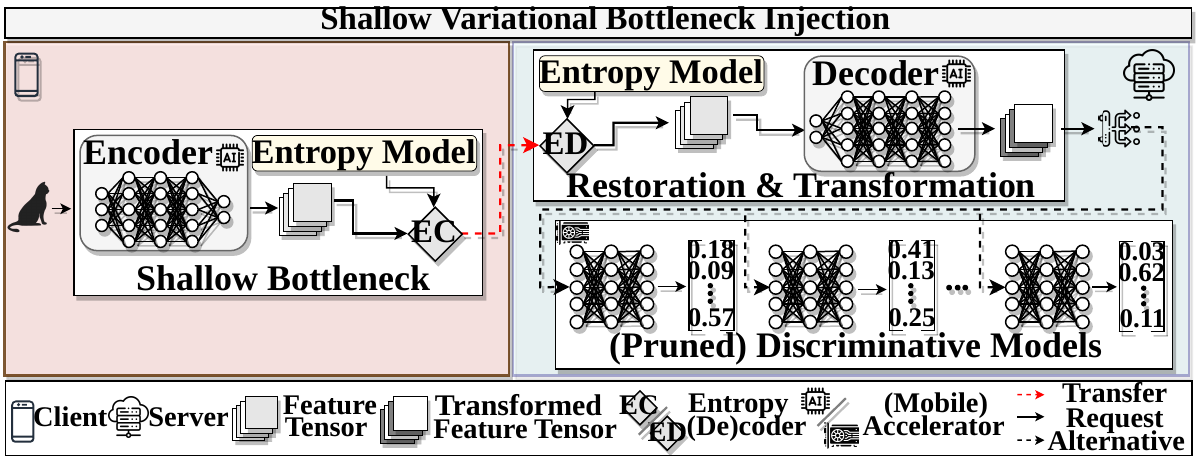}
    \caption{Prediction with Variational Bottleneck Injection}
    \label{fig:infstrategiesppt2}
\end{figure}
In \Cref{fig:infstrategiesppt2}, we sketch predictions with our proposed approach. There are two underlying distinctions to common SC methods. 

First, the model is not split between the client and the server. Instead, it deploys a lightweight encoder, and a decoder replaces the shallow layers of a backbone, i.e., the backbone is split within the server. A single decoder architecture corresponds to backbones with related architectures. Notably, a decoder restores and transforms the compressed signals to a backbone that may accommodate multiple tasks. The encoder is decoupled from a particular task and the decoder-backbone pair.
\Cref{subsec:networkarch} elaborates how separating the concerns permits one encoder instance to accommodate multiple decoder-backbone pairs.

Second, compared to split runtimes, the decision to apply the compression model may only depend on \textit{internal conditions}. It can decouple the client from any external component (e.g., server, router). Ideally, applying the encoder should always be preferable if a mobile device has the minimal required resources. Since our method does not alter the backbones, we do not need to maintain additional models to accommodate clients who cannot apply the encoder. Instead, we can simply route the image tensor to the input layer of the (unmodified) model.


The following describes the limitations of existing work for constrained devices to conceive a method with the abovementioned description.
\section{Problem Formulation} \label{sec:probform}
\begin{figure}[htb]
    \centering
\includegraphics[width=\columnwidth]{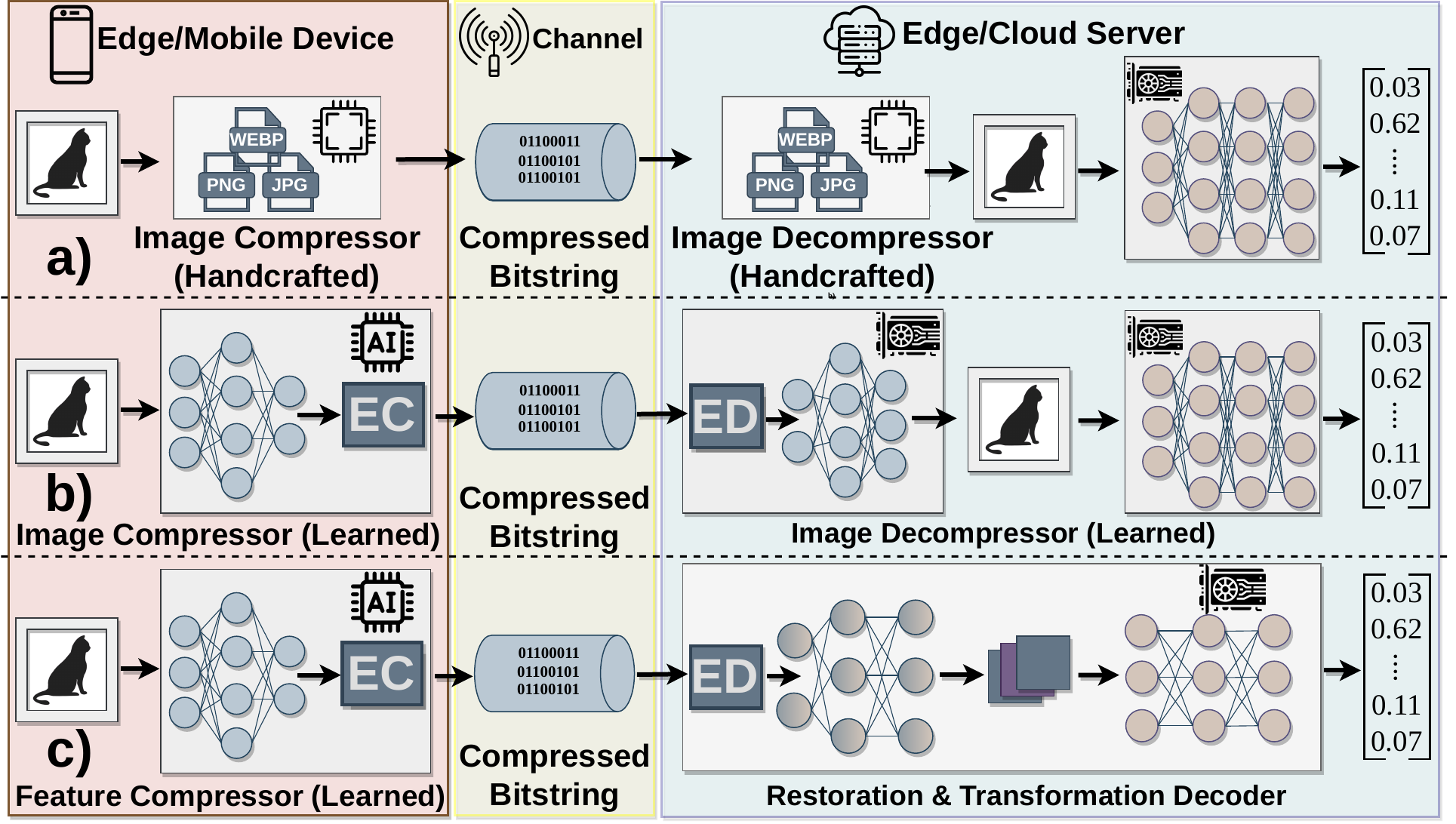}
    \caption{Utilizing client resources with (learned) codecs}
    \label{fig:compstrategies}
\end{figure}

The goal is for constrained clients to request real-time predictions from a large DNN while maximizing resource efficiency and minimizing bandwidth consumption with compression methods. \Cref{fig:compstrategies} illustrates the possible approaches when dedicating client resources exclusively for compression. 
Strategy a) corresponds to offloading strategies with CPU-bound handcrafted codecs. Strategy b) represents recent LIC models. Learned methods can achieve considerably lower bitrates with comparable distortion than commonly used handcrafted codecs~\cite{nonlineartc2020}. Nevertheless, we must consider that the overhead of executing large DNNs may dominate the reduced transfer time. 
Strategy c) is our advocated method with an embeddable variational feature compression that draws from the same underlying Nonlinear Transform Coding (NTC) framework as b). The challenge is to reduce overhead to make variational compression models suitable for real-time prediction with limited client resources.

To overcome the limitations of existing methods, we require
\begin{enumerate*}[label=(\roman*)] \item a resource-conscious encoder design. The encoder should minimize the transfer cost without increasing the predictive loss. 
Additionally, \item the decoder should exploit the available server-side resources without incurring significant overhead. Lastly, \item a compression model should fit for different downstream tasks and architectural families (e.g., CNNs or Vision Transformers)\end{enumerate*}.  


Before we can conceive an adequate method, we must formalize the properties of a suitable objective and elaborate on the limitations of existing methods when applied to shallow bottlenecks.
\subsection{Rate-Distortion Theory for Model Prediction}\label{subsec:taskspecrd}
By Shannon’s rate-distortion (r-d) theory~\cite{shannon1959coding}, we seek a mapping bound by a distortion constraint from a random variable (r.v.) $X$ to an r.v. $U$, minimizing the bitrate of the outcomes of $X$. More formally, given a distortion measure $\mathcal{D}$ and a distortion constraint $D_c$, the minimal bitrate is:
\begin{equation}
\underset{P_{U|X}}{\mathrm{min}}\; I(X;U)\; \text{s.t.}\; \mathcal{D}(X,U) \leq D_c\label{eq:rdbasics}
\end{equation}
where $I(X;U)$ is the mutual information and is defined as
\begin{equation}
\mathrm{I}(X;U)=\int_{}\int_{} p(x,u) \log \left(\frac{p(x,u)}{p(x) p(u)}\right) dxdu\label{eq:mutualinf}
\end{equation}
In lossy image compression, $U$ is typically the reconstruction of $\tilde{X}$ of the original input, and the distortion measure is some sum of squared errors $d(x, \tilde{x})$. Since the r-d theory does not restrict us to image reconstruction~\cite{berger1968rate}, we can apply distortion measures relevant to M2M communication. Notably, when our objective is to minimize predictive loss rather than reconstructing the input, we keep information that may be excessive for model predictions.

To elaborate on the potential of discarding information for discriminative tasks, consider the Data Processing Inequality (DPI). For any 3 r.v.s $X, Y, Z$ that form a Markov chain $X \leftrightarrow Y \leftrightarrow Z$ where the following holds:
\begin{equation}
I(X;Y) \geq I(X;Z)\label{eq:dpi}
\end{equation}
Then, describe the information flow in an $n$-layered sequential DNN, layer with the information path by viewing layered neural networks as a Markov chain of successive representations~\cite{tishby2015deep}:
\begin{equation}
I(X;Y) \geq I(R_1;Y) \geq I(R_2;Y) \geq \dots I(R_n;Y) \geq I(\tilde{Y}; Y)\label{eq:infopath}
\end{equation}
In other words, the final representation before a prediction $R_n$ cannot have more mutual information with the target than the input $X$ and typically has less.  
In particular, for high-level vision tasks that map a high dimensional input vector with strong pixel dependencies to a small set of labels, we can expect $I(X;Y) \gg I(\tilde{R_n}, Y)$.   
\subsection{From Deep to Shallow Bottlenecks}\label{subsec:fromdeeptoshallow}
When the task is to predict the ground-truth labels $Y$ from a joint distribution $P_{X,Y}$, the r-d objective is essentially given by the information bottleneck principle~\cite{informationbottleneck}. By relaxing the \eqref{eq:rdbasics} with a lagrangian multiplier, the objective is to maximize:
\begin{equation}
I(Z;Y) - \beta I(Z;X)\label{eq:vanillavib}
\end{equation}
Specifically, an encoding $Z$ should be a minimal sufficient statistic of $X$ respective $Y$, i.e., we want $Z$ to contain relevant information regarding $Y$ while discarding irrelevant information from $X$. Practical implementations differ by the target task and how they approximate \eqref{eq:vanillavib}. For example, an approximation of $I(Z;Y)$ for an arbitrary classification task the conditional cross entropy (CE)~\cite{introneuralcompression}:
\begin{equation}
 \mathcal{D} = H(P_Y, P_{\tilde{Y}|Z})\label{eq:distcondce} 
\end{equation}
Using \eqref{eq:distcondce} for estimating $I(Z;Y)$ to end-to-end optimize a neural compression model is not a novel idea (\Cref{subsubsec:featurecompression}). However, a common assumption in such work is that the latent variable is the final representation $R_n$ of a large backbone, which we refer to as \textit{Deep Variational Information Bottleneck Injection (DVBI)}. Conversely, we work with resource-constrained clients, i.e., to conceive lightweight encoders, we must shift the bottleneck to the shallow layers, which we refer to as \textit{Shallow Variational Bottleneck Injection (SVBI)}.
Intuitively, the existing methods for DVBI should generalize to SVBI, e.g., estimate the distortion term with \eqref{eq:distcondce} as in \cite{singh2020}. 

While shifting the bottleneck to the shallow layers results in an encoder with less capacity, the objective still approximates to \eqref{eq:rdbasics}. Yet, as we will show in \Cref{subsubsec:naivetydo}, applying the objective from \cite{singh2020} will result in incomparably worse results when moving the bottleneck to the shallow bottlenecks. 

A more promising method to estimate $I(Z;Y)$ is Head Distillation (HD)~\cite{matsubarahnd2019, cde2021} since it naturally aligns with shallow bottlenecks. As we will show in \Cref{subsec:rdperfevall}, HD  yields significantly better results than applying \eqref{eq:distcondce}. Surprisingly, despite showing promising results, HD is a suboptimal estimation for $I(Z;X)$ to approximate \cref{eq:rdbasics}. 

The following elaborates on SVBI and formulates the VIB objective for HD. 
\subsection{Head Distilled Deep Variational IB}\label{subsec:partialdistillvib}
Ideally, the bottleneck is embeddable in an existing predictor $P_{\mathcal{T}}$ without decreasing the performance. Therefore, it is not the hard labels $Y$ that define the task but the soft labels $Y_{\mathcal{T}}$. For simplicity, we handle the case for one task and defer discussion on multiple downstream tasks and DNNs to \Cref{subsec:networkarch}. 

To perform SVBI, take a copy of $P_{\mathcal{T}}$. Then, mark the location of the bottleneck by separating the copy into a head $\mathcal{P}_{h}$ and a tail $\mathcal{P}_{t}$. Importantly, both parts are deterministic, i.e., for every realization of r.v. $X$ there is a representation $\mathcal{P}_{h}(x) = h$ such that $\mathcal{P}_{\mathcal{T}}(x) = \mathcal{P}_{h}(\mathcal{P}_{t}(x))$.  Lastly, replace the head with an autoencoder and a parametric entropy model. 

The encoder is deployed at the sender, the decoder at the receiver, and the entropy model is shared. 
We distinguish between two optimization strategies to train the bottleneck’s compression model. First, is \textit{direct optimization} corresponding to the DVIB objective in \eqref{eq:vanillavib}, except we replace the CE with the standard KD loss~\cite{origdistill} to estimate $I(Z;Y)$. The second is \textit{indirect optimization} and describes HD with the objective:
\begin{equation}
I(Z;H) - \beta \; I(Z;X)\label{eq:vibdistilnaivel}
\end{equation}
Unlike the former, the latter does not directly correspond to \eqref{eq:rdbasics} for a representation $Z$ that is a minimal sufficient statistic of $X$ respective $Y_{\mathcal{T}}$. Instead, it replaces $Y$ with a proxy task for the compression model to replicate the output of the replaced head, i.e., training methods approximating \eqref{eq:vibdistilnaivel} optimize for a $Z$ that is a minimal sufficient statistic of $X$ respective $H$.
\begin{figure}[htb]
    \centering
\includegraphics[width=\columnwidth]{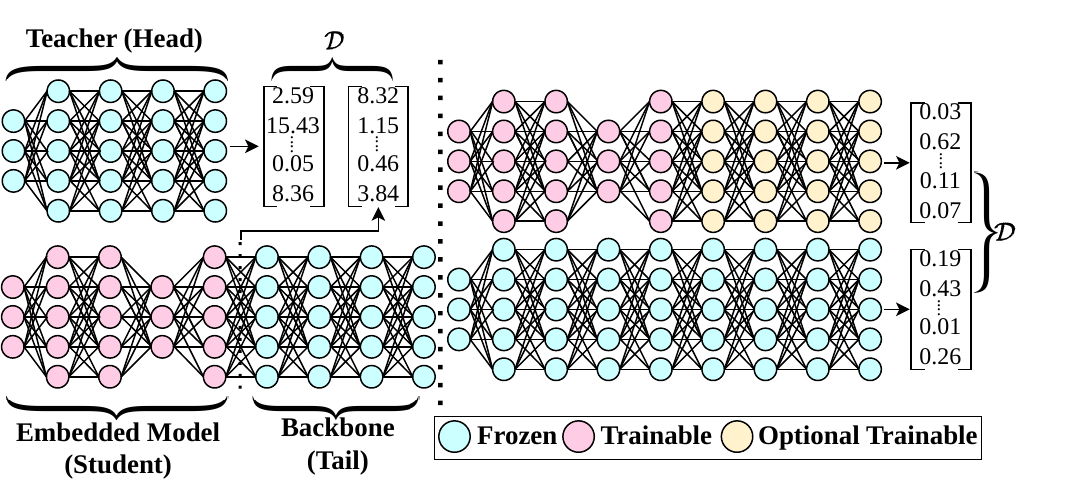}
    \caption{Left: Head Distillation. Right: Direct Optimization}
    \label{fig:trainstrategies}
\end{figure}
\Cref{fig:trainstrategies} illustrates the difference between estimating the objectives \eqref{eq:vanillavib} and \eqref{eq:vibdistilnaivel}.

With faithful replication of $H$, the partially modified DNN has an information path equivalent to its unmodified version. A sufficient statistic retains the information necessary to replicate the input for a deterministic tail, i.e., the final prediction does not change. The problem of \eqref{eq:vibdistilnaivel} is that it is a suboptimal approximation of \eqref{eq:rdbasics}. Although sufficiency still holds, it optimizes $Z$ respective $H$ and not $Y_{\mathcal{T}}$ to be minimal.

Based on the above formulations, the following proposes a practical method to train stochastic feature compression models. Additionally, it addresses the limitations of HD and includes architectural considerations. 
\section{Solution Approach} \label{sec:solapproach}
Our solution focuses on two distinct but intertwined aspects. First is an appropriate training objective. The second concerns a practical implementation by introducing an architectural design heuristic to accommodate backbones with various architectures with a single encoder architecture.
\subsection{Loss Function for End-to-end Optimization}\label{subsec:endtoendopt} 
We follow NTC~\cite{nonlineartc2020} to implement a neural compression algorithm. Specifically, we embed a stochastic compression model that we jointly optimize with an entropy model. 

Our objective resembles variational image compression optimization, as introduced in \cite{balle2017, balle2018}. For an image vector $x$, we have a parametric analysis transform $g_a(x; \phi_g)$ that maps $x$ to a latent vector $z$. Then, a quantizer $Q$ discretizes $z$ to $\bar{z}$, such that an entropy coder can use the entropy model to losslessly compress $\bar{z}$ to a sequence of bits. In learned image compression, a parametric synthesis transforms $g_{s}(\bar{z};\theta_{g})$ maps $\bar{z}$ to a reconstruction of the input $\tilde{x}$

However, we favor HD over direct optimization as a distortion measure since the former yields considerably better results even with a suboptimal loss function (\Cref{subsubsec:naivetydo}). Therefore, we require a $g_{s}(\bar{z};\theta_{g})$ that maps $\bar{z}$ to an approximation of a representation $\tilde{h}$ (i.e., the output of shallow layers of an arbitrary backbone). 

Analogous to variational inference, we approximate the intractable posterior $p(\tilde{z}|x)$ with a parametric variational density $q(\tilde{z}|x)$ as follows (excluding constants):
\begin{equation}
\mathbb{E}_{\boldsymbol{x} \sim p_{\boldsymbol{x}}} D_{\mathrm{KL}}\left[q \| p_{\tilde{\boldsymbol{z}} \mid \boldsymbol{x}}\right]=\mathbb{E}_{\boldsymbol{x} \sim p_{\boldsymbol{x}}} \mathbb{E}_{\tilde{\boldsymbol{z}} \sim q}[\underbrace{-\mathrm{log}\;p(x|\tilde{z})}_{\text{ distortion}} -\overbrace{\mathrm{log}\; p(\tilde{z})}^{\text{weighted rate}}]\label{eq:rdtermfp}
\end{equation}
By assuming a Gaussian distribution such that the likelihood of the distortion term is given by
\begin{equation}
P_{x|\tilde{z}}(x\; |\; \tilde{z}, \theta_g) = \mathcal{N}(x\; |\; g_s(\tilde{z};\theta_g), \;\mathbf{1})\label{eq:llasgaussian}
\end{equation}
we can use the square sum of differences between $h$ and $\tilde{h}$ as our distortion loss. 

The rate term describes the cost of compressing $\tilde{z}$. Analogous to the LIC methods discussed in \Cref{subsec:neuralcompression}, we apply uniform quantization $\bar{z} =  \round{\tilde{z}}$. Since discretization leads to problems with the gradient flow, we apply a continuous relaxation by adding uniform noise $\eta \thicksim \mathcal{U}(-\frac{1}{2}, \frac{1}{2})$. Combining the rate and distortion term, we derive the loss function for estimating objective \eqref{eq:vibdistilnaivel} as:
\begin{equation}
\mathcal{L}_{\mathrm{}} = \|\mathcal{P}_h(x) \text{ - } (g_s(g_a(x;\phi_g) + \eta;\theta_g)\|^2_2 + \beta \; \mathrm{log}(g_a(x;\theta_g) + \eta) \label{eq:rdlossbase}
\end{equation}
As described in \Cref{subsec:partialdistillvib}, by using HD for the distortion term, we rely on $H$ as a proxy target, i.e., the loss in \Cref{eq:rdlossbase} is a suboptimal approximation of \eqref{eq:rdbasics}.

The suboptimality stems from treating every pixel in $H$ equally important.
The implication here is that the MSE in \eqref{eq:rdlossbase} overly strictly penalizes pixels at spatial locations that contain redundant information that later layers can safely discard. Contrarily, the loss may not penalize the salient pixels enough when $\tilde{h}$ is numerically close to $h$. 

Hence, we can improve the loss in \eqref{eq:rdlossbase}
by introducing additional signals that regularize the suboptimal distortion term. The challenge is finding a tractable method that emphasizes the salient pixels necessary for multiple instances of a high-level vision task (e.g., classification of various datasets and labels). Moreover, the method should exclusively concern the loss function, i.e., it should not introduce any additional model components or operations during inference.
\subsection{Saliency Guided Distortion} \label{subsec:saliencymaps}
We consider HD an extreme form of Hint Training (HT)~\cite{wang2021knowledge, romero2014fitnets} where the hint becomes the primary objective rather than an auxiliary regularization term. Sbai et al. perform deterministic bottleneck injection with HD using the suboptimal distortion term~\cite{cde2021}. Nevertheless, their method only considers dimensionality reduction without a parametric entropy model as an approximation to compression, i.e., it is generalized by the loss in \eqref{eq:rdbasics}$(\beta=0$). 
Matsubara et al. add further hints from the deeper layers by extending the distortion term with the sum of squared between the deeper layers~\cite{matsubarahnd2019, matsubara2022supervised}. This approach has several downsides besides prolonged train time. The distortion term may now dominate the rate term, i.e., without exhaustively tuning the hyperparameters for each distortion term, the optimization algorithm should favor converging towards local optima. Moreover, we show \Cref{subsubsec:compes} that pure HD can significantly outperform this method using the loss in \Cref{eq:rdbasics} without the hints from the deeper layers.

In principle, we could improve the performance by extracting signals from deeper layers and directly transferring them to the bottleneck. The caveat is that the effectiveness of knowledge distillation decreases for teachers when the student has considerably less capacity than the teacher~\cite{wang2021knowledge}. Hence, instead of directly introducing hints at the encoder, we propose regularizing the distortion term with saliency maps.

For each sample, we require a vector $\mathcal{S}$, where each $s_i \in \mathcal{S}$ is a weight term for a spatial location salient about the conditional probability distributions of the remaining tail layers. Then, we should be able to improve the r-d performance by regularizing the distortion term in \eqref{eq:rdlossbase} with
\begin{equation}
\mathcal{L}_{\mathrm{distortion}} = 
\gamma_{1} \cdot \mathcal{L}_1 + \gamma_{2} \cdot s_i  \cdot \frac{1}{N}\sum_{i}(h_i - \tilde{h}_i)^2 \label{eq:salientldistortion}
\end{equation}
Where $\mathcal{L}_1$ is the distortiont term from \Cref{eq:rdlossbase}, and $\gamma_{1}, \gamma_{2}$ are nonnegative real numbers summing to $1$. We default to 
$\gamma_{1} = \gamma_{2} = \frac{1}{2}$ in our experiments. 
\begin{figure}[htb]
    \centering
    \includegraphics[width=\columnwidth]{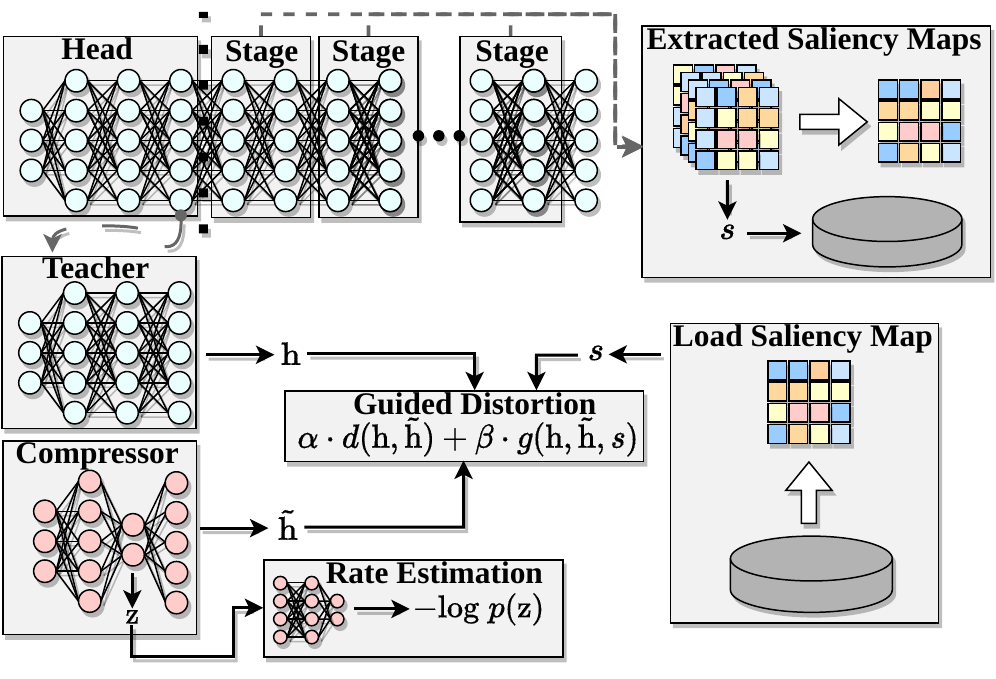}
    \caption{Training setup}
    \label{fig:trainsetup}
\end{figure}
\Cref{fig:trainsetup} describes our final training setup. Note that we only require computing the saliency maps once, and they are architecturally agnostic towards the encoder.

\begin{figure}[htb]
    \centering
    \includegraphics[width=\columnwidth]{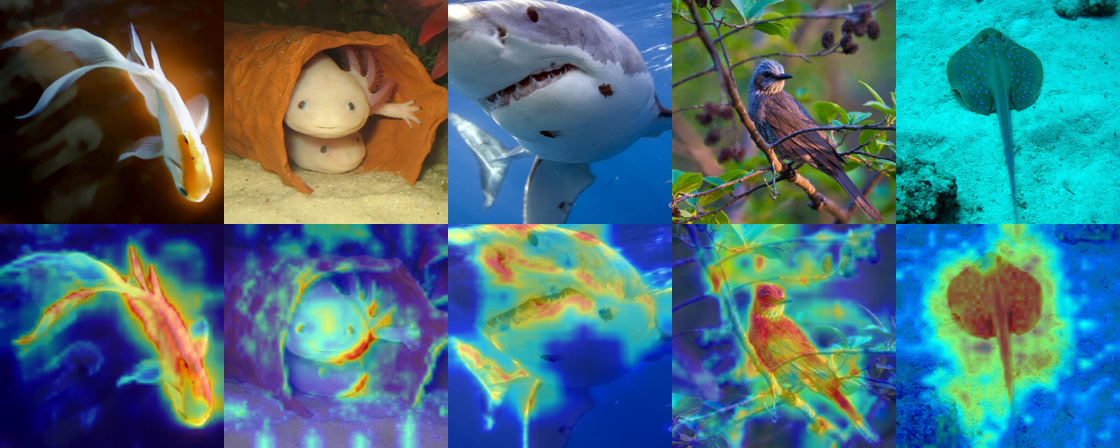}
    \caption{Extracted saliency maps using Grad-CAM}
    \label{fig:gradcam}
\end{figure}

We derive the saliency maps using \textit{class activation mapping (CAM)}~\cite{camorig}. Although CAMs are typically used to improve the explainability of DNNs, they suit our purposes by allowing us to summarize salient pixel locations.
Specifically, we use a variant of Grad-CAM~\cite{gradcam} to measure a spatial location’s importance at any stage. \Cref{fig:gradcam} illustrates some examples of saliency maps when averaged over the deeper backbone stages.
In this work, we favor Grad-CAM over (more intricate) methods due to its architecture-agnostic nature and computational efficiency. For example, mixing with guided backpropagation~\cite{guidedbackprop} could refine the resulting saliency maps with finer-grained feature importance scaling. However, guided backpropagation relies on specific properties of the activation function and requires adjustments for each architectural family. 
\subsection{Network Architecture} \label{subsec:networkarch}
The beginning of this section broke down our aim into three problems. We addressed the first with SVBI and proposed a novel training method for low-capacity compression models. A generalizable resource-asymmetry-aware autoencoder design remains. Additionally, the encoder should be reusable for several backbones. To not inflate the significance of our contribution, we refrain from including components based on existing work in efficient neural network design. 
\subsubsection{Model Taxonomy} \label{subsubsec:modeltax}
\begin{figure}[ht]
    \centering
    \includegraphics[width=\columnwidth]{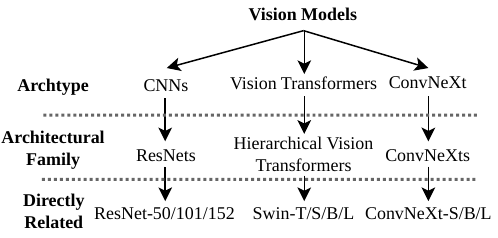}
    \caption{Simple taxonomy with minimal example}
    \label{fig:taxonomyexample}
\end{figure}

We introduce a minimal taxonomy described in \Cref{fig:taxonomyexample} for our approach.
The top-level, \textit{Archtype}, reflects the primary inductive bias of the model. \textit{Architectural families} describe variants (e.g., ResNets such as ResNet~\cite{resnetorig}, Wide ResNet~\cite{wideresnet}, ResNeXt~\cite{resnext}, etc.). \textit{Directly related} refers to the same architecture of different sizes (e.g., Swin-T, Swin-S, Swin-B, etc.).
The challenge is to conceive a design heuristic that can exploit the available server resources to aid the lightweight encoder with minimal overhead on the prediction task.
First, we concretize shallow features by describing how to locate the layers for bottleneck placement. 
Then, we derive the heuristic to conceive decoder models for arbitrary architectural families and how to account for client-server resource asymmetry.
    
Lastly, we describe how to share trained compressor components among directly related architectures. 
\subsubsection{Bottleneck Location by Stage Depth} \label{subsubsec:bloctax}
Consider how most modern DNNs consist of an initial embedding followed by a few stages (Described in \Cref{subsec:limsc}). Within directly related architectures, the individual components are identical. The difference between variants is primarily the embed dimensions or the block ratio of the deepest stage. For example, the block ratio of ResNet-50 is 3:4:\textbf{6}:3, while the block ratio of ResNet-101 is 3:4:\textbf{23}:3. Consequently, the stage-wise organization of models defines a natural interface for SVBI. For the remainder of this work, we refer to the \textit{shallow layers} as the layers before the deepest stage (i.e., the initial embedding and the first two stages).

\subsubsection{Decoder Blueprints} \label{subsubsec:decoderblueprint}
A key characteristic distinguishing archetypes is the inductive bias introduced by basic building blocks (e.g., convolutions versus attention layers). To consider the varying representations among non-related architectures, we should not disregard architecture-induced bias by directly repurposing neural compression models for SC. For example, a scaled-down version of Ballé et al.’s~\cite{balle2017} convolutional neural compression model can yield strong r-d performance for bottlenecks reconstructing a convolutional layer~\cite{sc2bench}.
However, we will show that this does not generalize to other architectural families, such as hierarchical vision transformers~\cite{swin2021}.

One potential solution is to use identical components for the compression model from a target network. While this may be inconsequential for server-side decoders,
it is inadequate for encoders due to the heterogeneity of edge devices. Vendors have varying support for the basic building blocks of a DNN, and particular operations may be prohibitively expensive for the client. 
Hence, in FrankenSplit, the encoder is fixed, but the decoder is adaptable. Regardless of the decoder architecture, we account for the heterogeneity with a uniform encoder architecture composed of three downsampling residual blocks of two stacked $3 \times 3$ convolutions with ReLU non-linearity, totaling around $140,000$ parameters. We handle the varying representations by introducing \textit{decoder blueprints} tailored towards an architectural family, i.e., one blueprint corresponds to all directly related architectures. 
\begin{figure}[ht]
    \centering
\includegraphics[width=\columnwidth]{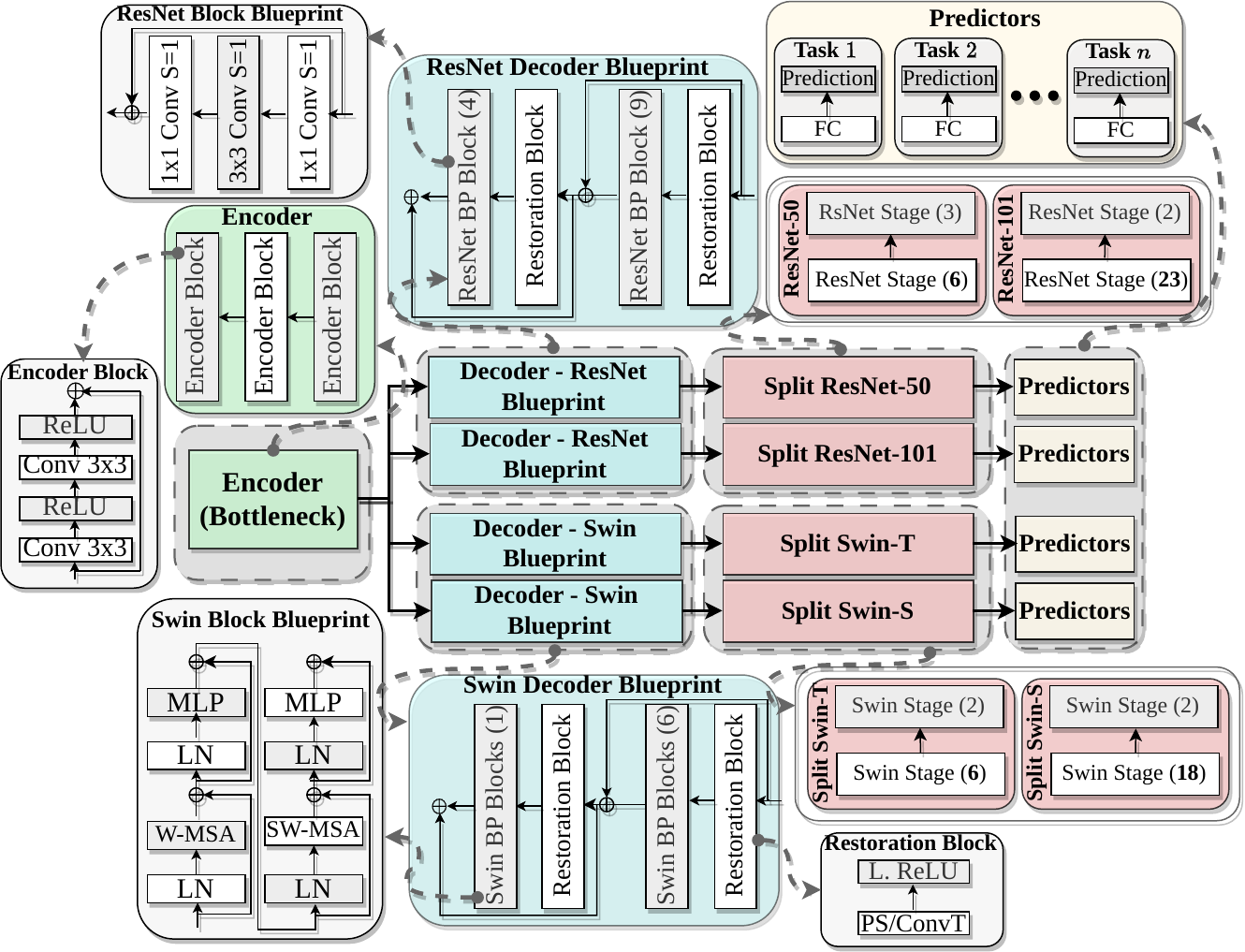}
    \caption{Reference implementation of FrankenSplit}
    \label{fig:referencarch}
\end{figure}

\Cref{fig:referencarch} illustrates a reference implementation of FrankenSplit post-training with two blueprints applied to two variants. Creating blueprints is required only once for an architectural family. 
Boxes within the gray areas are separate instances (i.e., only one encoder), and boxes with the same name share an architecture. The rounded boxes outside organize layer views from coarse to fine-grained. 
We elaborate on how a single encoder can accommodate multiple decoder-backbone pairs in \Cref{subsubsec:encreusabile}. The numbers in the parentheses refer to stage depth. Since the backbones are foundational models extensively trained on large datasets, we can naturally accommodate several downstream tasks by attaching separately trained predictors. 

Blueprint instances replace a backbone's first two stages (i.e., the shallow layers) with two blueprint stages, taking a compressed representation as input instead of the original sample.  The work by Liang et al.~\cite{swinir2021} inspires our approach to treat decoding as a restoration problem. Each stage comprises a restoration block and several blueprint (transformation) blocks, followed by a residual connection. The idea is to separate restoration (i.e., upsampling, ‘’smoothing`` quantized features) from transformation (i.e., matching the target representation regardless of encoder architecture). The restoration block is agnostic regarding the target architecture and optionally upsamples. The blueprint blocks induce the same bias as the target architectural family. 

Two distinctions exist between the original blocks and their corresponding blueprint (transformation). First, the latter modifies operations not to reduce the latent spatial dimensions. 
Second, the embedding layer dimensions and stage depths may differ to reflect the resource asymmetry commonly found in MEC.


Although we should consider the resource asymmetry between the client and the server (i.e., by allocating more parameters to the decoder), there are limitations. Learning a function that can accurately retain necessary information is limited by the encoder's capacity (\Cref{subsec:taskspecrd}). Still, when end-to-end optimizing the compression model, it can benefit from increasing the decoder’s capacity for restoration with diminishing returns.

Intuitively, we implement blueprints that result in decoder instances with, at most, the same execution time as the head of a target backbone. As a reminder, unlike most work in SC, we advocate keeping the execution time roughly equal on the server rather than reducing it. The encoder’s responsibility is not to minimize the server load by executing shallow backbone layers. FrankenSplit treats the encoder entirely separate from the backbone. Besides dedicating the encoder exclusively to reducing transfer size, this separation of concern is necessary to accommodate several backbones with a single encoder instance.
\subsubsection{Encoder Re-Usability} \label{subsubsec:encreusabile}
We argue that the representation of shallow layers generalizes well enough that it is possible to reuse compressor components. 
\begin{figure}[ht]
    \centering
    \includegraphics[width=\columnwidth]{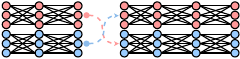}
    \caption{Routing head outputs to different tails}
    \label{fig:rerouteexp}
\end{figure}
Consider the experiment illustrated in \Cref{fig:rerouteexp}, where we split several backbones into head and tail models. The backbones are off-the-shelf models from \textit{torch image models} (timm)~\cite{rw2019timm} and pre-trained on the ImageNet~\cite{imagenet} dataset.
The head models consist of the initial embedding and shallow layers, i.e., the first two stages. The remaining layers comprise the substantially larger tails (roughly ~$2-5\%$ of total model parameters).

Then, we freeze the tail parameters and route the head output to all non-corresponding tails (e.g., ConvNeXt-T to Swin-T/S/B) and measure the accuracy every few iterations with a batch size of 128 as we finetune the head parameters using cross entropy loss. Each head-tail pair is a separate model built by attaching a copy of the head from one architecture to the tail of another. Where dimensions between head and tail pairs do not match, we add a single $1\times1$ convolutional layer. 
\begin{figure}[ht]
    \centering
    \includegraphics[width=\columnwidth]{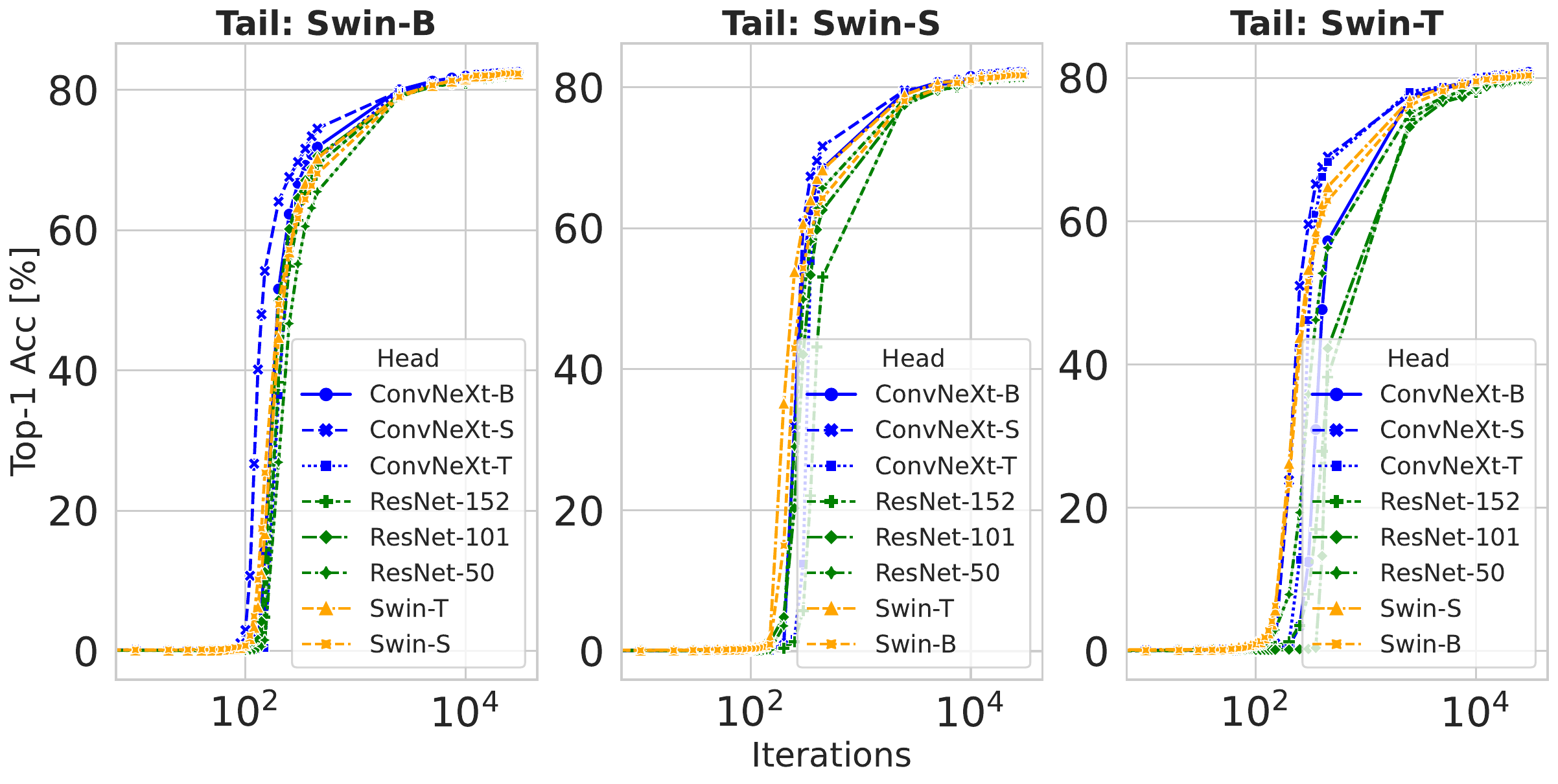}
    \caption{Recovering Top-1  accuracy of rerouted heads}
    \label{fig:headintrawithft}
\end{figure}

\Cref{fig:headintrawithft} shows how rerouting the input between head models first (0 iterations) results in near 0\% accuracy across all head-tail pairs. However, the concatenated models quickly converge near their original accuracy (roughly $80-83\%$) within just a few iterations ($10100$ iterations with 128 samples corresponding to one epoch on the ImagNet dataset). Notice that this holds regardless of whether the head-tail pairs are directly related to the modified network.

Therefore, if a compressor can sufficiently approximate the representation of just one head (i.e., the shallow layers of a network), it should be possible to accommodate arbitrary tails (i.e., the deeper layers of a network). 

Crucially, applying the distortion measure in \eqref{eq:rdlossbase} or \eqref{eq:salientldistortion} does not result in an inherently different encoder behavior. Like training the compression model with a distortion measure from LIC, the purpose of the encoder is reducing uncertainty by decorrelating the data and discarding information. The distortion measure only controls what information an encoder should prioritize. Regardless of the target backbone’s architecture, the encoder should decorrelate the input to reduce uncertainty. Conversely, the decoder seeks a mapping to the backbone’s representation. 

In other words, if we can map the latent to one representation, we can map it to any other with comparable information content. We can freeze the encoder and train various decoders to support arbitrary architectures once we train one compression model with a particular teacher as described in \Cref{fig:trainsetup}. 
The blueprints facilitate an efficient transformation from the encoder’s compressed representation to an input suitable for a particular backbone. 




Notice that this method keeps the encoder parameters frozen, permitting us to deploy a single set of weights across all clients. Moreover, it does not modify the backbones at any step. After deployment, splitting is replaced with
rerouting the input to a layer index (\Cref{subsubsec:bloctax}). Then, we can serve clients with the same models regardless of whether they applied the compressor.

\section{Evaluation}\label{sec:eval}
\subsection{Training \& Implementation Details}\label{subsec:impl}
We optimize our compression models initially on the 1.28 million ImageNet~\cite{imagenet} training samples for 15 epochs, as described in \cref{subsec:endtoendopt} and \cref{subsec:saliencymaps}, with some slight practical modifications for stable training. We aim to minimize bitrate without sacrificing predictive strength. Hence, we first seek the lowest $\beta$ resulting in lossless prediction.

We use Adam optimization~\cite{kingma2014adam} with a batch size of $16$ and start with an initial learning rate of $1 \cdot 10^{-3}$, then gradually lower it to $1 \cdot 10^{-6}$ with an exponential scheduler.

To implement our method, we use PyTorch~\cite{pytorch2018pytorch}, CompressAI~\cite{begaint2020compressai} for entropy estimation and entropy coding, and pre-trained backbones from timm~\cite{rw2019timm}. All baseline implementations and weights were either taken from CompressAI or the official repository of a baseline. To compute the saliency maps, we use a modified \textit{XGradCAM} method from the library in~\cite{jacobgilpytorchcam} and include necessary patches in our repository. Lastly, to ensure reproducibility, we use torchdistill~\cite{matsubara2021torchdistill}. 
\subsection{Experiment Setting} \label{subsec:blineandexp}
The experiments reflect the deployment strategies illustrated in \Cref{fig:compstrategies} and \Cref{fig:infstrategiesppt2}. Ultimately, we must evaluate whether FrankenSplit enables latency-sensitive and performance-critical applications. Regardless of the particular task, a mobile edge client requires access to a DNN with high predictive strength on a server. Therefore, we must show whether FrankenSplit adequately solves two problems associated with offloading high-dimensional image data for real-time discriminative tasks. First, whether it considerably reduces the bandwidth consumption compared to existing methods without sacrificing predictive strength. Second, whether it improves inference times over various communication channels, i.e., it must remain competitive even when stronger connections are available.

Lastly, the evaluation should assess whether our method generalizes to arbitrary backbones. However, since it is infeasible to perform exhaustive experiments on all existing visual models, we focus on three well-known representatives and a subset of their variants instead.
Namely, (i) ResNet~\cite{resnetorig} for classic residual CNNs. (ii) Swin Transformer~\cite{swin2021} for hierarchical vision transformers, which are receiving increasing adaptation for a wide variety of vision tasks. (iii) ConvNeXt~\cite{liu2022convnet} for modernized state-of-the-art CNNs. \Cref{tab:bboverview} summarizes the relevant characteristics of the unmodified backbones subject to our experiments.
\begin{table}[htb]
\centering
\caption{Overview of Backbone Performance on Server}
\label{tab:bboverview}
\resizebox{\columnwidth}{!}{%
\begin{tabular}{@{}ccccc@{}}
\toprule
Backbone & Ratios & Params & \begin{tabular}[c]{@{}c@{}}Inference\\ (ms)\end{tabular} & \begin{tabular}[c]{@{}c@{}}Top-1 Acc.\\ (\%)\end{tabular} \\ \midrule
Swin-T     & 2:2:6:2  & 28.33M & 4.77  & 81.93 \\
\rowcolor[HTML]{EDEDED} 
Swin-S     & 2:2:18:2 & 49.74M & 8.95  & 83.46 \\
Swin-B     & 2:2:30:2 & 71.13M & 13.14 & 83.88 \\ \midrule
ConvNeXt-T & 3:3:9:3  & 28.59M & 5.12  & 82.70 \\
\rowcolor[HTML]{EDEDED} 
ConvNeXt-S & 3:3:27:3 & 50.22M & 5.65  & 83.71 \\
ConvNeXt-B & 3:3:27:3 & 88.59M & 6.09  & 84.43 \\ \midrule
ResNet-50  & 3:4:6:3  & 25.56M & 5.17  & 80.10 \\
\rowcolor[HTML]{EDEDED} 
ResNet-101 & 3:4:23:3 & 44.55M & 10.17 & 81.91 \\
ResNet-152 & 3:8:36:3 & 60.19M & 15.18 & 82.54 \\ \bottomrule
\end{tabular}%
}
\end{table}
\subsubsection{Baselines} \label{subsubsec:baselines} Since our work aligns closest to learned image compression, we extensively compare FrankenSplit with learned and handcrafted codecs applied to the input images, i.e., the input to the backbone is the distorted output. Comparing task-specific methods to general-purpose image compression methods may seem unfair. However, FrankenSplit’s universal encoder has up to 260x less trainable parameters and further reduces overhead by not including side information or a sequential context model. 

The naming convention for the learned baselines is the first author's name, followed by the entropy model. Specifically, we choose the work by Ballé et al.~\cite{balle2017, balle2018} and Minnen et al.~\cite{minnen2018} for LIC methods since they represent foundational milestones. Complementary, we include the work by Cheng et al.~\cite{cheng2020} to demonstrate improvements with architectural enhancement. 

As the representative for disregarding autoencoder size to achieve state-of-the-art r-d performance in LIC, we chose the work by Chen et al.~\cite{chen2021} Their method differs from other LIC baselines by using a partially parallelizable context model, which trades off compression rate with execution time according to the configurable block size. We refer to such context models as Blocked Joint Hierarchical Priors and  Autoregressive (BJHAP). Due to the large autoencoder, we found evaluating the inference time on constrained devices impractical when the context model is purely sequential and set the block size to 64x64. Additionally, we include the work by Lu et al.~\cite{lu2022} as a milestone of the recent effort on efficient LIC with reduced autoencoders but only for latency-related experiments since we do not have access to the trained weights. 

As a baseline for the state-of-the-art SC, we include the Entropic Student (ES)~\cite{sc2bench, matsubara2022supervised}. The ES demonstrates the performance of directly applying a minimally adjusted LIC method for feature compression.
One caveat is that we intend to show how FrankenSplit generalizes beyond CNN backbones, despite the encoder’s simplistic CNN architecture. Although Matsubara et al. evaluate the ES on a wide range of backbones, most have no lossless configurations. Nevertheless, comparing bottleneck injection methods using different backbones is fair, as we found that the choice does not significantly impact the r-d performance (\Cref{subsubsec:diffbb}). Therefore, for an intuitive comparison, we choose ES with ResNet-50 using the same factorized prior entropy model as FrankenSplit. 

We separate the experiments into two categories to assess whether our proposed method addresses the abovementioned problems. 
\subsubsection{Criteria rate-distortion performance} 
We measure the bitrate in bits per pixel (bpp) because it permits directly comparing models with different input sizes.  Choosing a distortion measure to draw meaningful and honest comparisons is challenging for feature compression. 

Unlike evaluating reconstruction fidelity for image compression, PSNR or MS-SSIM does not provide intuitive results regarding predictive strength. Similarly, reporting absolute values (e.g., top-1 accuracy) gives an unfair advantage to experiments conducted on higher capacity backbones and veils the efficacy of a proposed method. 

Hence, for a transparent evaluation, we determine the adversarial effects of codecs with image classification since it provides an unambiguous performance metric with established benchmark datasets. Specifically, we evaluate the distortion with the relative measure \textit{predictive loss}, i.e., the drop in top-1 accuracy incurred by codecs. In particular, for SVBI methods, (near) lossless prediction implies that the reconstruction is a sufficient approximation for shallow features of an arbitrary feature extractor.

To ensure a fair comparison, we give the LIC and handcrafted baselines a grace threshold of 1.0\% top-1 accuracy, to account for mitigating predictive loss incurred by codec artifacts~\cite{luo2020rate}. For FrankenSplit, we set the threshold at 0.4\%, reflecting the configuration with the lowest predictive loss of the ES. 
Note that, unlike the ES, FrankenSplit does not rely on finetuning the tail parameters of a backbone to improve r-d performance.
\subsubsection{Measuring latency and overhead}
To account for the resource asymmetry in MEC, we use NVIDIA Jetson boards\footnote{nvidia.com/en-gb/autonomous-machines/embedded-systems/} for representing capable but resource-constrained mobile clients. Contrastingly, the server hosts a powerful GPU. 
\Cref{tab:hwconf} summarizes the hardware we use in our experiments.
\begin{table}[htb]
\centering
\caption{Clients and Server Hardware Configuration}
\label{tab:hwconf}
\resizebox{\columnwidth}{!}{%
\begin{tabular}{cccc}
\hline
Device       & Arch    & CPU                 & GPU         \\ \hline
Server       & x86     & 16x Ryzen @ 3.4 GHz & RTX 3090    \\
\rowcolor[HTML]{EDEDED} 
Client (TX2) & arm64x8 & 4x Cortex @ 2 GHz   & Vol. 48 TC  \\
Client (NX)  & arm64x8 & 4x Cortex @ 2 GHz   & Pas. 256 CC \\ \hline
\end{tabular}%
}
\end{table}

\subsection{Rate-Distortion Performance} \label{subsec:rdperfevall}
\begin{figure}[htb]
    \centering
    \includegraphics[width=\columnwidth]{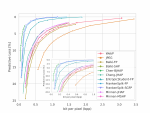}
    \caption{Rate-distortion curve for ImageNet}
    \label{fig:rdcurveilsvrc}
\end{figure}

We measure the predictive loss by the drop in top-1 accuracy from \Cref{tab:bboverview} using the ImageNet validation set for the standard classification task with 1000 categories. Analogously, we measure filesizes of the entropy-coded binaries to calculate the average bpp. To demonstrate that we can accommodate a non-CNN backbone with a CNN encoder, we start with a Swin-B implementation of FrankenSplit. \Cref{fig:rdcurveilsvrc} shows r-d curves with the Swin-B backbone.  
The architecture of FrankenSplit-FP (FS-FP) and FrankenSplit-SGFP (FS-SGFP) are identical. We train both models with the loss functions derived in \Cref{subsec:endtoendopt}. The difference is that FS-SGFP is saliency-guided, i.e., FS-FP represents the pure HD training method and is an ablation to the saliency-guided distortion.
\subsubsection{Effect of Saliency Guidance}
Although FS-FP performs better than almost all other models, it is trained with the suboptimal objective discussed in \Cref{subsec:partialdistillvib}. We identified the issue as overly skewing the objective needlessly towards the distortion term. Consequently, we proposed regularizing the distortion term by applying extracted saliency maps in \Cref{subsec:saliencymaps} to improve the r-d performance. We favor Grad-CAM to compute the saliency maps over comparable methods for two reasons. First, it is generically applicable to arbitrary vision models. Second, it does not introduce additional tunable hyperparameters. 
The suboptimality of the unregularized objective is demonstrated by FS-SGFP outperforming FS-FP. By simply guiding the distortion loss with saliency maps, we achieve a 25\% lower bitrate without impacting predictive strength or additional runtime overhead.  
\subsubsection{Comparison to the ES} \label{subsubsec:compes}
Even without saliency guidance, FS-FP consistently outperforms ES by a large margin. Specifically, FS-FP and FS-SGFP achieve 32\% and 63\% lower bitrates for the lossless configuration. 

We ensured that our bottleneck injection incurs comparable overhead for a direct comparison to the ES. Moreover, the ES has an advantage due to finetuning tail parameters in an auxiliary training stage. Therefore, we attribute the performance gain to the more sophisticated architectural design decisions described in \Cref{subsec:networkarch}.
\subsubsection{Comparison to Image Codecs}
For almost all lossy codec baselines, \Cref{fig:rdcurveilsvrc} illustrates that FS-(SG)FP has a significantly better r-d performance. 
Comparing FS-FP to Ballé-FP demonstrates the r-d gain of task-specific compression over general-purpose image compression. Although the encoder of FrankenSplit has 25x fewer parameters, both codecs use an FP entropy model with encoders consisting of convolutional layers. Yet, the average file size of FS-FP with a predictive loss of around 5\% is 7x less than the average file size of Ballé-FP with comparable predictive loss. 

FrankenSplit also beats modern general-purpose LIC without including any of their 
heavy-weight components. The only baseline FrankenSplit does not convincingly outperform is Chen-BJHAP. Nevertheless, in \Cref{subsec:advlearned}, we demonstrate that the incurred overhead offsets the compression gain disproportionately.
\subsubsection{Image Codec Incurred Predictive Loss} 
For clarity, we separately evaluate r-d performance on the other backbones listed in \Cref{tab:bboverview} for FrankenSplit and baseline codecs.

\begin{figure}[htb]
    \centering
    \includegraphics[width=\columnwidth]{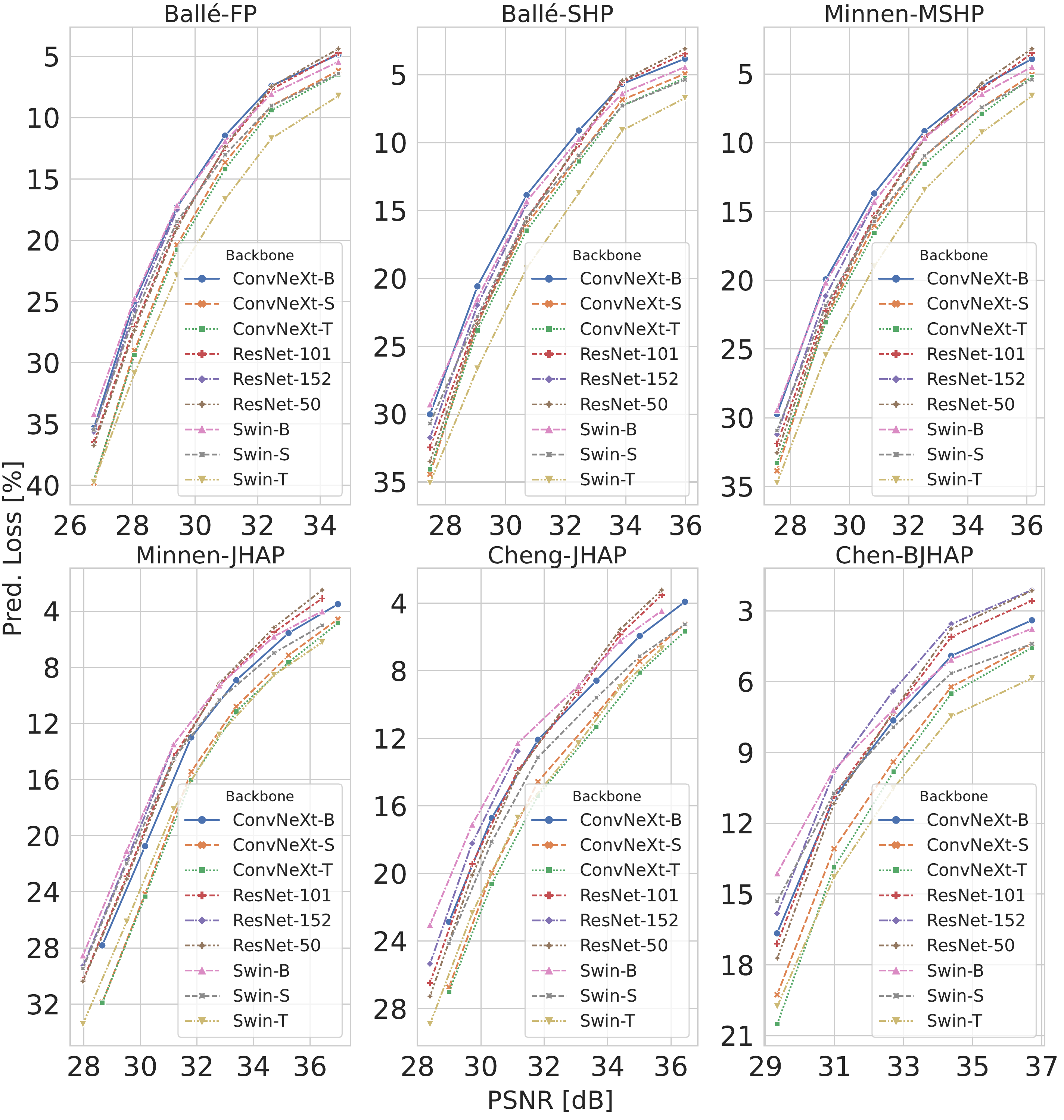}
    \caption{Predictive Loss of baselines on multiple Backbones}
    \label{fig:psnrvsacc}
\end{figure}

Earlier, we argued that measuring PSNR is unsuitable to assess effects on downstream prediction. 
Since the image codecs are entirely decoupled from the predictive task, the bitrate is identical regardless of the backbone. We use this opportunity to plot PSNR instead of bpp against predictive loss in \Cref{fig:psnrvsacc}. 

Considering that compression models aggressively discard information, it is intuitive that the predictive loss is comparable across backbones. While some models handle distorted samples better, the difference in predictive loss is at most 3-5\%. Still, the discrepancy demonstrates that PSNR is not a suitable measure for downstream tasks even within the same codec. More importantly, the discrepancy across baselines is considerably wider. For example, it is around 10\% between Minnen-MSHP and Chen-BJHAP for lower PSNR levels.
\subsubsection{Blueprints Generalization to Arbitrary Backbones} \label{subsubsec:diffbb}
We now evaluate the r-d performance of other implementations of FrankenSplit to determine whether the blueprint heuristics generalize to arbitrary architectures.  We create a decoder blueprint  (\Cref{subsubsec:decoderblueprint}) for each of the three architectural families (Swin, ResNet, and ConvNeXt). Then, we perform bottleneck injection at the layers before the deepest stage (\Cref{subsubsec:bloctax}),  
\begin{figure}[htb]
    \centering
    \includegraphics[width=\columnwidth]{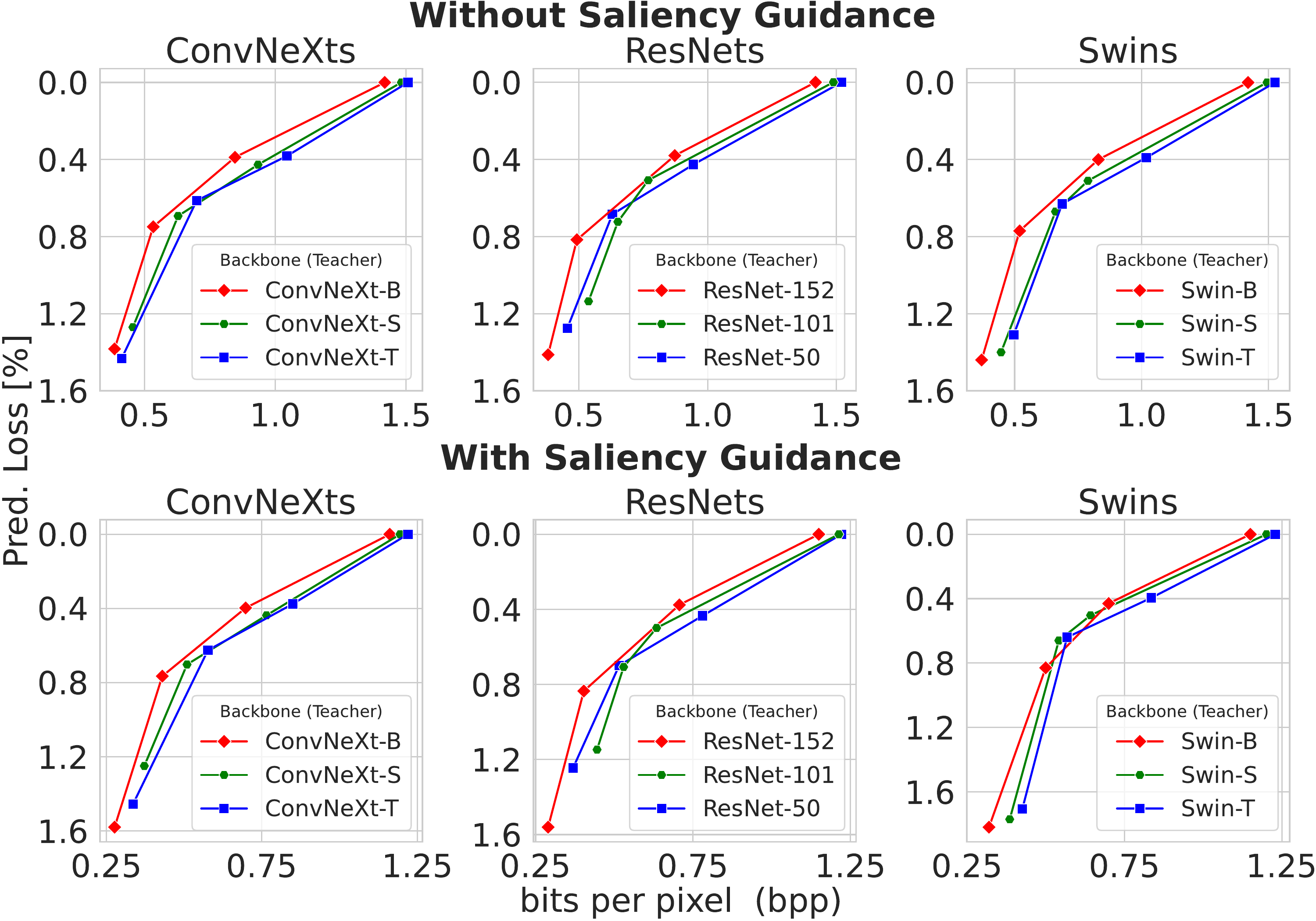}
    \caption{Rate-distortion curve for various backbones}	
    \label{fig:rdplotsallbbs}
\end{figure}
\Cref{fig:rdplotsallbbs} plots r-d performance of directly related architectures sharing the corresponding blueprint but with separately trained compressors. All models are trained as described in \Cref{fig:trainsetup}. Across all architectural families, we observe similar r-d performance. The (near) lossless configurations of the largest backbones (Swin-B,  ConvNeXt-B, ResNet-152) require around the same bpp, whereas smaller models tend to require 3-4\% more bpp for comparable predictive loss. 

Next, we conduct experiments to determine the importance of finding an adequate blueprint but assigning mismatching instances to a backbone.
\Cref{tab:bismismatch} summarizes the results for the largest backbones with varying decoder sizes. The Swin blueprint for the Swin-B decoder results in the FrankenSplit implementation from FS-FP from \Cref{fig:rdcurveilsvrc}. With 1\% overhead in parameters, the compressor achieves 5.08 kB for $0.40\%$ predictive loss. However, once we train compressors with ResNet or ConvNeXt restoration blocks, the r-d performance for the Swin-B is significantly worse when overhead is roughly equal.
\begin{table}[htb]
\centering
\caption{Effect of Mismatching Blueprints }
\label{tab:bismismatch}
\resizebox{\columnwidth}{!}{%
\begin{tabular}{cccc}
\hline
Blueprint-Backbone              & Params Overhead (\%)   & File Size (kB)                & Pred. Loss (\%)              \\ \hline
                                &                        & \cellcolor[HTML]{EDEDED}19.05 & \cellcolor[HTML]{EDEDED}2.49 \\
                                & \multirow{-2}{*}{0.96} & 15.07                         & 3.00                         \\
                                &                        & \cellcolor[HTML]{EDEDED}14.46 & \cellcolor[HTML]{EDEDED}2.53 \\
                                & \multirow{-2}{*}{2.86} & 11.37                         & 2.74                         \\
                                &                        & \cellcolor[HTML]{EDEDED}12.53 & \cellcolor[HTML]{EDEDED}1.36 \\
\multirow{-6}{*}{ConvNext-Swin} & \multirow{-2}{*}{5.89} & 10.08                         & 2.09                         \\ \hline
                                &                        & \cellcolor[HTML]{EDEDED}22.54 & \cellcolor[HTML]{EDEDED}0.82 \\
                                & \multirow{-2}{*}{1.03} & 18.19                         & 0.99                         \\
                                &                        & \cellcolor[HTML]{EDEDED}16.32 & \cellcolor[HTML]{EDEDED}0.81 \\
                                & \multirow{-2}{*}{2.73} & 12.68                         & 0.98                         \\
                                &                        & \cellcolor[HTML]{EDEDED}13.89 & \cellcolor[HTML]{EDEDED}0.79 \\
\multirow{-6}{*}{ResNet-Swim}   & \multirow{-2}{*}{5.25} & 10.01                         & 0.98                        
\end{tabular}%
}
\end{table}
A blueprint that performs well for its intended target architecture results in substantially worse r-d performance for other architecture. 
Only increasing the decoder size brings the r-d performance closer to configurations that apply the appropriate blueprint. 

From our findings, we can draw several conclusions. The r-d performance regarding the backbone network is near-agnostic. The implication is that the information content of the teachers (i.e., shallow layers) of varying architectures is comparable. We explain this by considering that we select the shallow layers as all layers preceding the deepest stage, which have comparable parameters across varying architectures.

Additionally, choosing a decoder architecture with the correct inductive bias (i.e., a blueprint) can transform compressed features significantly more efficiently.
\subsubsection{Single Encoder with Multiple Backbones} \label{subsubsec:multenc}
We conduct a similar experiment as head rerouting from \Cref{subsubsec:encreusabile}. However, we finetune the decoders instead of the head models.

We first select the compressors with (a near) lossless prediction from \Cref{fig:rdplotsallbbs} for each architectural family. Then, we choose the encoder from one of the compressors corresponding to the largest variants.
Finally, we attach the decoders from the other compressors and finetune their parameters. We use unweighted head distillation and cross entropy (between the backbone classifier outputs and the hard labels) as the loss function. Analogous to the experiment in \Cref{subsubsec:encreusabile}, we set the batch size as 128 and use PyTorch's Adam optimizer with a learning rate of $7 \cdot 10^{-5}$. 
\begin{figure}[htb]
    \centering
    \includegraphics[width=\columnwidth]{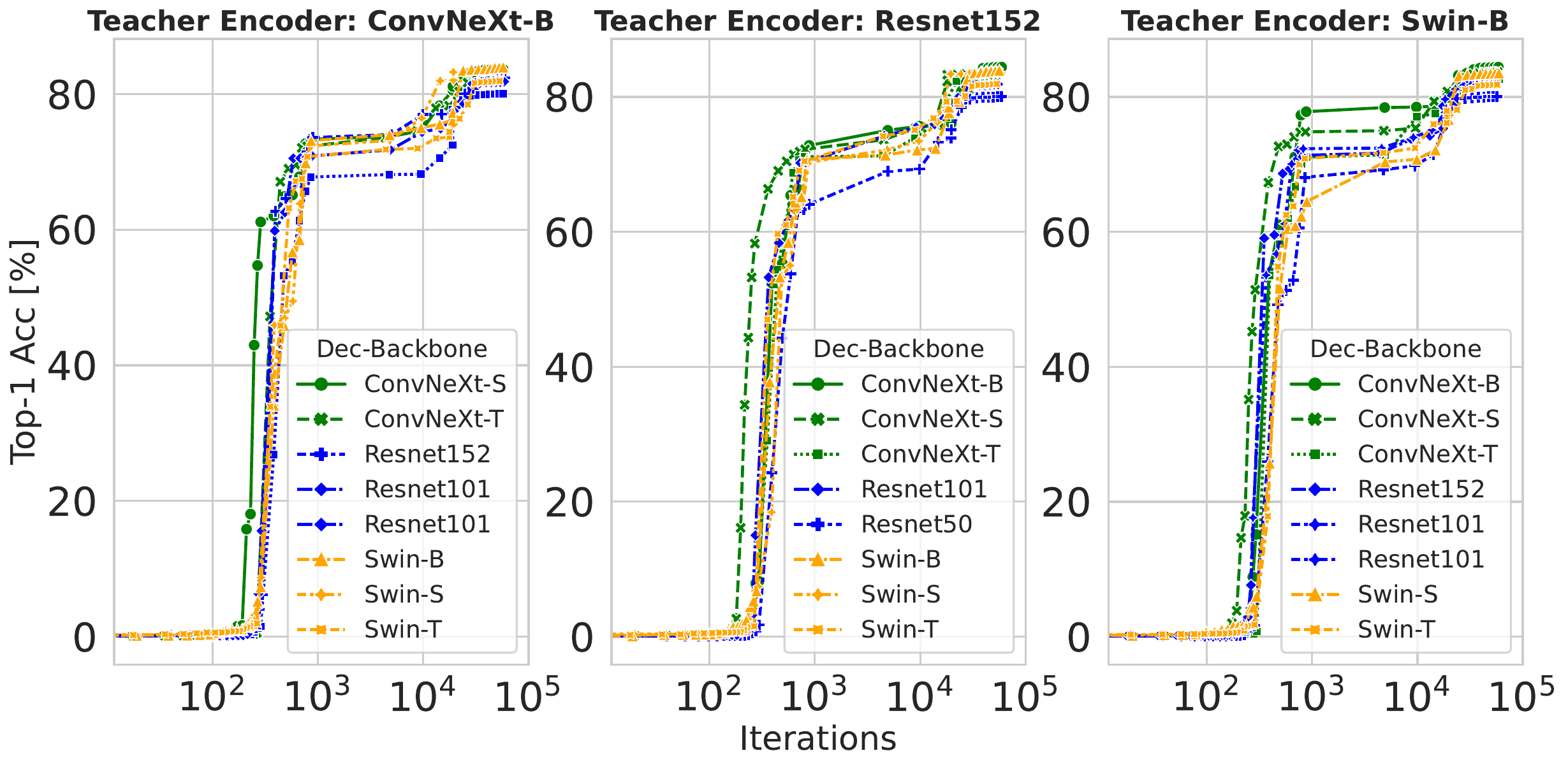}
    \caption{Iterations to recover accuracy with decoder}
    \label{fig:plotsfintrabneck}
\end{figure}
To demonstrate the limited importance of the initial teacher, we repeat this process for each of the three encoders separately and summarize the results in \Cref{fig:plotsfintrabneck}. 
Note that the bitrate does not change due to freezing the encoder parameters. Hence, we report iterations until accuracy is restored to exemplify the similarity to the rerouting experiment in \Cref{subsubsec:encreusabile}. We consider an accuracy restored if it is within $0.25 \pm 0.25\%$ of its original accuracy.

Besides requiring more iterations for convergence, the results are  unsurprisingly similar to the head routing experiment outlined in \Cref{fig:headintrawithft}. Since we can infer from the earlier results that decoders can sufficiently approximate the head output, finetuning the decoder is near-equivalent to finetuning a head. 
\subsubsection{Generalization to multiple Downstream Tasks} \label{subsubsec:genmultask}
Arguably, SVBI naturally generalizes to multiple downstream tasks due to approximating shallow features.
We provide empirical evidence by evaluating the r-d performance of the compressors from \Cref{fig:rdcurveilsvrc} without retraining the weights on different datasets.

Specifically, attach separate classifiers to the Swin-B backbone (as illustrated in \Cref{fig:referencarch}). Using PyTorch's Adam optimizer, we train each classifier for five epochs with no augmentation, a learning rate of $5 \cdot 10^{-5}$. A classifier refers to the last layers of a network.

\begin{figure*}[htbp]
    \centering
    \includegraphics[width=\textwidth]{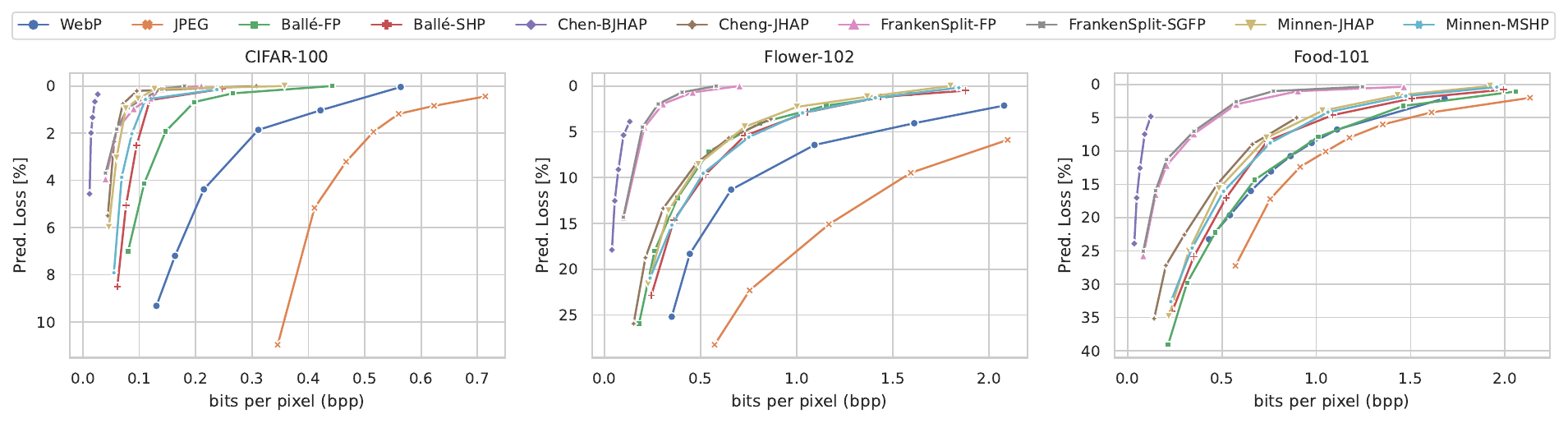}
    \caption{Rate-distortion curve for multiple downstream tasks}
    \label{fig:rdfinetune}
\end{figure*}

For FrankenSplit-(SG)FP, we applied none or only rudimentary augmentation to evaluate how our method handles a type of noise it did not encounter during training. Hence, we include the Food-101~\cite{bossard2014food} dataset since it contains noise in high pixel intensities. Additionally, we include CIFAR-100~\cite{krizhevsky2009learning}. Lastly, we include Flower-102~\cite{flowers102} datasets to contrast more challenging tasks. The classifiers achieve an  $87.73\%$, $88.01\%$, and $89.00\%$ top-1 accuracy, respectively. \Cref{fig:rdfinetune} summarizes the r-d curves for each task.
Our method still demonstrates clear r-d performance gains over the baselines. More importantly, notice how FS-SGFP outperforms FS-FP on the r-d curve for the Food-101 dataset, with a comparable margin to the ImageNet dataset. Contrarily, on the Flower-102 datasets, there is less performance difference. Presumably, on simple datasets, the suboptimality of HD is less significant. Considering how easier tasks require less model capacity, the diminishing efficacy saliency guidance is consistent with our claims from \Cref{sec:probform}. 
\subsubsection{Effect of Tensor Dimensionality on R-D Performance} \label{subsubsec:tensordimrdperf}
\Cref{subsec:ftdimquant} argues that measuring tensor dimensionality is inadequate to assess whether partial execution on the client is worthwhile.

\begin{figure}[htbp]
    \centering
    \includegraphics[width=\columnwidth]{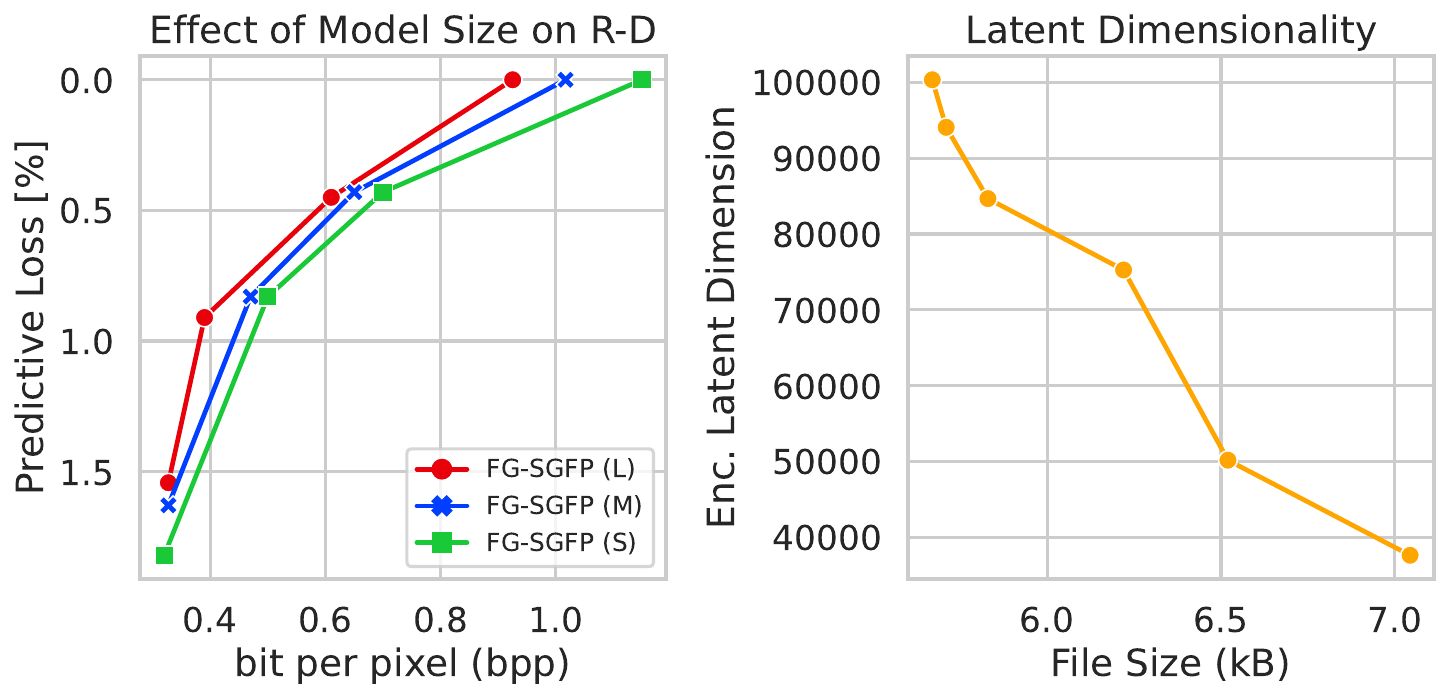}
    \caption{Comparing effects on sizes}
    \label{fig:effectsizes}
\end{figure}

To verify, we implement and train additional instances of FrankenSplit with the Swin-B backbone and show results in \Cref{fig:effectsizes}
FS-SGFP(S) is the model with a small encoder ($\sim$$140'000$ parameters) we have used for our previous results. FS-SGFP(M) and FS-SGFP(L) are medium and large models where we increased the (output) channels $C=48$ to $96$ and $128$, respectively. Besides the number of channels, we’ve trained the medium and large models using the same configurations. On the left, we plot the r-d curves showing that increasing encode capacity naturally results in lower bitrates without additional predictive loss. For the plot on the right, we train further models with $C = \{48, 64,  96, 108,  120,  128 \}$ using the configuration resulting in lossless prediction. Notice how increasing output channels will result in higher dimensional latent tensors $C \times 28 \times 28$ but inversely correlates to compressed file size. Arguably, increasing the encoder capacity will yield more powerful transforms to decorrelate the input.
\subsubsection{The Limitations of Direct Optimization for SVBI} \label{subsubsec:naivetydo}
\begin{figure}[htbp]
    \centering
    \includegraphics[width=\columnwidth]{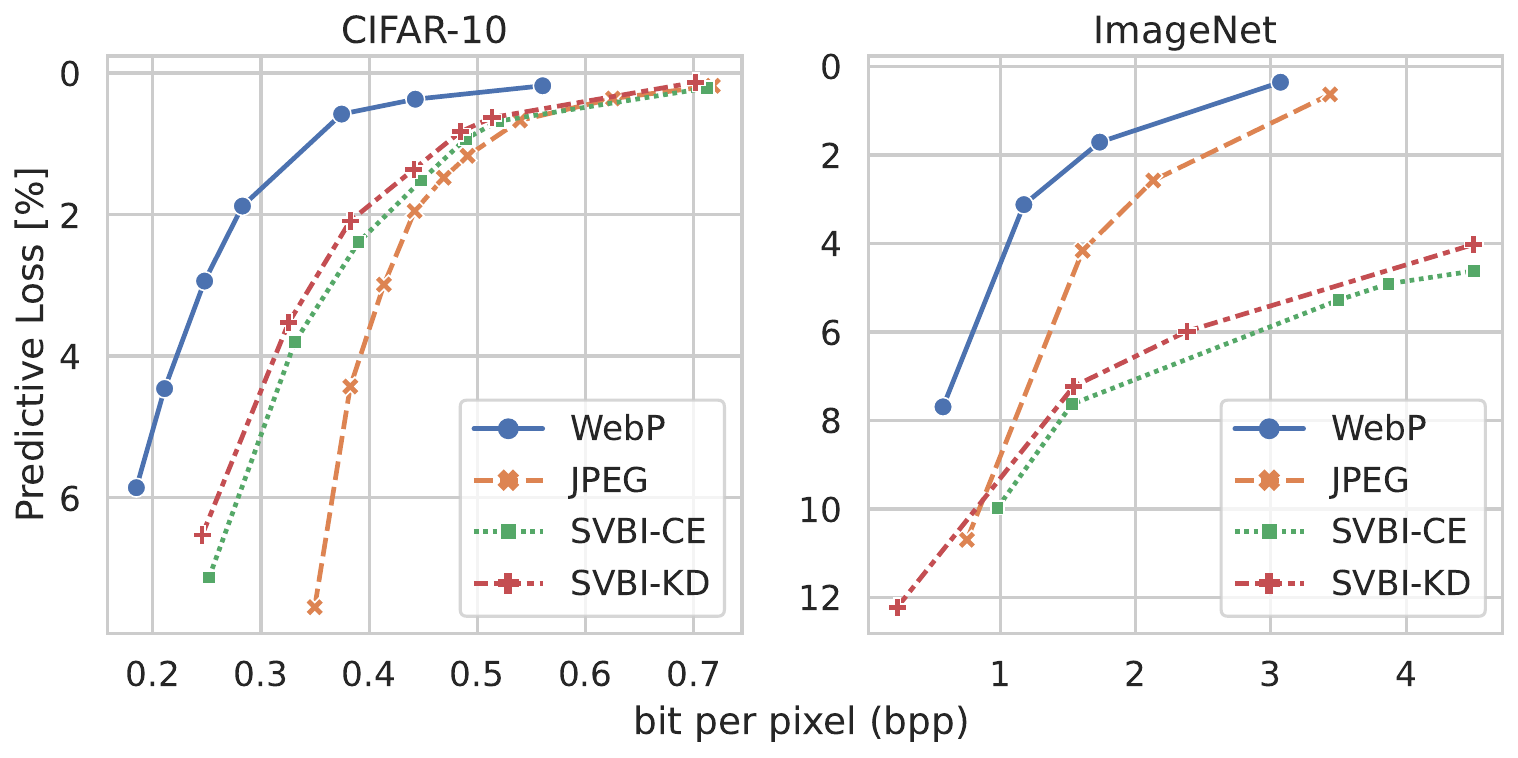}
    \caption{Contrasting the r-d performance}
    \label{fig:rdnaivety}
\end{figure}
Section \ref{subsec:endtoendopt} mentioned that direct optimization does not work for SVBI as it does for DVBI, where the bottleneck is at the penultimate layer. Specifically, it performs incomparably worse than HD despite the latter's inherent suboptimality. We demonstrate this by applying the SVBI-CE and SVBI-KD objective on the CIFAR-10~\cite{krizhevsky2009learning} and ImageNet dataset. All models are identical and trained with the setup in \Cref{subsec:impl}, except we train for more epochs to account for slower convergence. 

\Cref{fig:rdnaivety} summarizes the results the results. On CIFAR-10, SVBI-CE and SVBI-KD yield moderate performance gain over JPEG. Yet, they perform substantially worse on ImageNet. 

Sufficiency as a necessary precondition may explain why the objective in \eqref{eq:vanillavib} does not yield good results when the bottleneck is at a shallow layer, as the mutual information $I(Y;\tilde{Y})$ is not adequately high. Since the representation of the last hidden and shallow layer are so far apart in the information path, there is insufficient information to minimize $\mathcal{D}(H; \tilde{H})$. 
The compression model approximates the intermediate representation for a simple classification task to minimize predictive loss by incurring higher bitrates. Consequently, for the challenging ImageNet classification task, the same method incurs significant predictive loss even when skewing the r-d objective heavily towards high bitrates.

\subsection{Prediction Latency and Overhead} \label{subsec:latencyeval}
We exclude entropy coding from our measurement, since not all baselines use the same entropy coder. For brevity, the results implicitly assume the Swin-B backbone for the remainder of this section. Inference times with other backbones for FrankenSplit can be derived from \Cref{tab:exectimefs}. 
\begin{table}[htb]
\centering
\caption{Execution Times of FS (S) with Various Backbones}
\label{tab:exectimefs}
\resizebox{\columnwidth}{!}{%
\begin{tabular}{cccc}
\hline
Backbone &
  \begin{tabular}[c]{@{}c@{}}Overhead Prams\\ (\%)\end{tabular} &
  \begin{tabular}[c]{@{}c@{}}Inf. Server+NX \\ (m/s)\end{tabular} &
  \begin{tabular}[c]{@{}c@{}}Inf. Server+TX2\\ (m/s)\end{tabular} \\ \hline
Swin-T     & 2.51 & 7.83  & 9.75  \\
\rowcolor[HTML]{EDEDED} 
Swin-S     & 1.41 & 11.99 & 13.91 \\
Swin-B     & 1.00 & 16.12 & 18.04 \\ \hline
ConvNeXt-T & 3.46 & 6.83  & 8.75  \\
\rowcolor[HTML]{EDEDED} 
ConvNeXt-S & 1.97 & 8.50  & 10.41 \\
ConvNeXt-B & 0.90 & 9.70  & 11.62 \\ \hline
ResNet-50  & 3.50 & 13.16 & 10.05 \\
\rowcolor[HTML]{EDEDED} 
ResNet-101 & 2.01 & 8.13  & 15.08 \\
ResNet-152 & 1.48 & 18.86 & 20.78 \\ \hline
\end{tabular}%
}
\end{table}
Analogously, the inference times of applying LIC models for different unmodified backbones can be derived using \Cref{tab:bboverview}. Notably, the relative overhead decreases the larger the tail is, which is favorable since we target inference from more accurate predictors.
\subsubsection{Computational Overhead} \label{subsubsec:compoverhead}
We first disregard network conditions to get an overview of the computational overhead of applying compression models. \Cref{tab:infcompexecnobw} summarizes the execution times of the prediction pipeline’s components. 
\begin{table}[htb]
\centering
\caption{Inference Pipeline Components Execution Times}
\label{tab:infcompexecnobw}
\resizebox{\columnwidth}{!}{%
\begin{tabular}{ccccc}
\hline
Model &
  \begin{tabular}[c]{@{}c@{}}Prams \\ Enc./Dec.\end{tabular} &
  \begin{tabular}[c]{@{}c@{}}Enc. {[}NX/TX2{]}\\ (ms)\end{tabular} &
  \begin{tabular}[c]{@{}c@{}}Dec. \\ (ms)\end{tabular} &
  \begin{tabular}[c]{@{}c@{}}Full {[}NX/TX2{]}\\ (ms)\end{tabular} \\ \hline
FrankenSplit &
  \textbf{\begin{tabular}[c]{@{}c@{}}0.14M/\\ 2.06M\end{tabular}} &
  \textbf{\begin{tabular}[c]{@{}c@{}}2.92/\\ 4.87\end{tabular}} &
  2.00 &
  \textbf{\begin{tabular}[c]{@{}c@{}}16.34/\\ 18.29\end{tabular}} \\
\rowcolor[HTML]{EDEDED} 
Ballé-FP &
  \begin{tabular}[c]{@{}c@{}}3.51M/\\ 3.51M\end{tabular} &
  \begin{tabular}[c]{@{}c@{}}27.27/\\ 48.93\end{tabular} &
  \textbf{1.30} &
  \begin{tabular}[c]{@{}c@{}}41.71/\\ 63.37\end{tabular} \\
Ballé-SHP &
  \begin{tabular}[c]{@{}c@{}}8.30M/\\ 5.90M\end{tabular} &
  \begin{tabular}[c]{@{}c@{}}28.16/\\ 50.89\end{tabular} &
  1.51 &
  \begin{tabular}[c]{@{}c@{}}42.81/\\ 65.54\end{tabular} \\
\rowcolor[HTML]{EDEDED} 
Minnen-MSHP &
  \begin{tabular}[c]{@{}c@{}}14.04M/\\ 11.65M\end{tabular} &
  \begin{tabular}[c]{@{}c@{}}29.51/\\ 52.39\end{tabular} &
  1.52 &
  \begin{tabular}[c]{@{}c@{}}44.17/\\ 67.05\end{tabular} \\
Minnen-JHAP &
  \begin{tabular}[c]{@{}c@{}}21.99M/\\ 19.59M\end{tabular} &
  \begin{tabular}[c]{@{}c@{}}4128.17/\\ 4789.89\end{tabular} &
  275.18 &
  \begin{tabular}[c]{@{}c@{}}4416.7/\\ 5078.2\end{tabular} \\
\rowcolor[HTML]{EDEDED} 
Cheng-JHAP &
  \begin{tabular}[c]{@{}c@{}}16.35M/\\ 22.27M\end{tabular} &
  \begin{tabular}[c]{@{}c@{}}2167.34/\\ 4153.95\end{tabular} &
  277.26 &
  \begin{tabular}[c]{@{}c@{}}2457.7/\\ 4444.3\end{tabular} \\
Lu-JHAP &
  \begin{tabular}[c]{@{}c@{}}5.28M/\\ 4.37M\end{tabular} &
  \begin{tabular}[c]{@{}c@{}}2090.88/\\ 5011.56\end{tabular} &
  352.85 &
  \begin{tabular}[c]{@{}c@{}}2456.8/\\ 5377.8\end{tabular} \\
\rowcolor[HTML]{EDEDED} 
Chen-BJHAP &
  \begin{tabular}[c]{@{}c@{}}36.73M/\\ 28.08M\end{tabular} &
  \begin{tabular}[c]{@{}c@{}}3111.01/\\ 5837.38\end{tabular} &
  43.16 &
  \begin{tabular}[c]{@{}c@{}}3167.3/\\ 5893.6\end{tabular} \\ \hline
\end{tabular}%
}
\end{table}
Enc. NX/TX2 refers to the encoding time on the respective client device. Analogously, dec. refers to the decoding time at the server. Lastly, Full NX/TX2 is the total execution time of encoding at the respective client plus decoding and the prediction task at the server.
Lu-JHAP demonstrates how LIC models without a sequential context component are noticeably faster but are still 9.3x-9.6x slower than FrankenSplit despite a considerably worse r-d performance. 
Notice that the computational load of FrankenSplit is near evenly distributed between the client and the server.
The significance of considering resource asymmetry is emphasized by how the partially parallelized context model of Chen-BJHAP leads to faster decoding on the server. Nevertheless, it is slower than other JHAP baselines due to the overhead of the increased encoder size outweighing the performance gain of the blocked context model on constrained hardware. 
\subsubsection{Competing against Offloading} 
The average compressed filesize gives the transfer size from the ImageNet validation set. 
\begin{table}[htb]
\centering
\caption{Total Latency with Various Wireless Standards}
\label{tab:latencywithchannel}
\resizebox{\columnwidth}{!}{%
\begin{tabular}{ccccc}
\hline
\begin{tabular}[c]{@{}c@{}}Standard/\\ Data Rate\\ (Mbps)\end{tabular} &
  codec &
  \begin{tabular}[c]{@{}c@{}}Transfer\\ (ms)\end{tabular} &
  \begin{tabular}[c]{@{}c@{}}Total {[}TX2{]}\\ (ms)\end{tabular} &
  \begin{tabular}[c]{@{}c@{}}Total {[}NX{]}\\ (ms)\end{tabular} \\ \hline
 &
  FS-SGFP (0.23) &
  142.59 &
  160.48 &
  158.53 \\
 &
  \cellcolor[HTML]{EDEDED}FS-SGFP (LL) &
  \cellcolor[HTML]{EDEDED}209.89 &
  \cellcolor[HTML]{EDEDED}227.78 &
  \cellcolor[HTML]{EDEDED}225.83 \\
 &
  Minnen-MSHP &
  348.85 &
  415.89 &
  393.01 \\
 &
  \cellcolor[HTML]{EDEDED}Chen-BJHAP &
  \cellcolor[HTML]{EDEDED}40.0 &
  \cellcolor[HTML]{EDEDED}6167.79 &
  \cellcolor[HTML]{EDEDED}3441.41 \\
 &
  WebP &
  865.92 &
  879.06 &
  879.06 \\
\multirow{-6}{*}{\begin{tabular}[c]{@{}c@{}}BLE/\\ 0.27\end{tabular}} &
  \cellcolor[HTML]{EDEDED}PNG &
  \cellcolor[HTML]{EDEDED}2532.58 &
  \cellcolor[HTML]{EDEDED}2545.72 &
  \cellcolor[HTML]{EDEDED}2545.72 \\ \hline
 &
  FS-SGFP (0.23) &
  3.21 &
  21.09 &
  19.15 \\
 &
  \cellcolor[HTML]{EDEDED}FS-SGFP (LL) &
  \cellcolor[HTML]{EDEDED}4.72 &
  \cellcolor[HTML]{EDEDED}22.61 &
  \cellcolor[HTML]{EDEDED}20.66 \\
 &
  Minnen-MSHP &
  7.85 &
  74.89 &
  52.01 \\
 &
  \cellcolor[HTML]{EDEDED}Chen-BJHAP &
  \cellcolor[HTML]{EDEDED}0.9 &
  \cellcolor[HTML]{EDEDED}6128.69 &
  \cellcolor[HTML]{EDEDED}3402.31 \\
 &
  WebP &
  19.48 &
  32.63 &
  32.63 \\
\multirow{-6}{*}{\begin{tabular}[c]{@{}c@{}}4G/\\ 12.0\end{tabular}} &
  \cellcolor[HTML]{EDEDED}PNG &
  \cellcolor[HTML]{EDEDED}56.98 &
  \cellcolor[HTML]{EDEDED}70.13 &
  \cellcolor[HTML]{EDEDED}70.13 \\ \hline
 &
  FS-SGFP (0.23) &
  0.71 &
  18.6 &
  16.65 \\
 &
  \cellcolor[HTML]{EDEDED}FS-SGFP (LL) &
  \cellcolor[HTML]{EDEDED}1.05 &
  \cellcolor[HTML]{EDEDED}18.93 &
  \cellcolor[HTML]{EDEDED}16.99 \\
 &
  Minnen-MSHP &
  1.74 &
  68.78 &
  45.9 \\
 &
  \cellcolor[HTML]{EDEDED}Chen-BJHAP &
  \cellcolor[HTML]{EDEDED}0.2 &
  \cellcolor[HTML]{EDEDED}6127.99 &
  \cellcolor[HTML]{EDEDED}3401.61 \\
 &
  WebP &
  4.33 &
  17.47 &
  17.47 \\
\multirow{-6}{*}{\begin{tabular}[c]{@{}c@{}}Wi-Fi/\\ 54.0\end{tabular}} &
  \cellcolor[HTML]{EDEDED}PNG &
  \cellcolor[HTML]{EDEDED}12.66 &
  \cellcolor[HTML]{EDEDED}25.81 &
  \cellcolor[HTML]{EDEDED}25.81 \\ \hline
 &
  FS-SGFP (0.23) &
  0.58 &
  18.46 &
  16.51 \\
 &
  \cellcolor[HTML]{EDEDED}FS-SGFP (LL) &
  \cellcolor[HTML]{EDEDED}0.85 &
  \cellcolor[HTML]{EDEDED}18.73 &
  \cellcolor[HTML]{EDEDED}16.78 \\
 &
  Minnen-MSHP &
  1.41 &
  68.44 &
  45.56 \\
\multirow{-4}{*}{\begin{tabular}[c]{@{}c@{}}5G/\\ 66.9\end{tabular}} &
  \cellcolor[HTML]{EDEDED}Chen-BJHAP &
  \cellcolor[HTML]{EDEDED}0.16 &
  \cellcolor[HTML]{EDEDED}6127.95 &
  \cellcolor[HTML]{EDEDED}3401.57 \\
\multicolumn{1}{l}{} &
  WebP &
  3.49 &
  16.64 &
  16.64 \\
\multicolumn{1}{l}{} &
  \cellcolor[HTML]{EDEDED}PNG &
  \cellcolor[HTML]{EDEDED}10.22 &
  \cellcolor[HTML]{EDEDED}23.36 &
  \cellcolor[HTML]{EDEDED}23.36 \\ \hline
\end{tabular}%
}
\end{table}Using the transfer size, we evaluate transfer time on a broad range of standards. Since we did not include the execution time of entropy coding for learned methods, the encoding and decoding time for the handcrafted codecs is set to 0. 
The setting favors the baselines because both rely on sequential CPU-bound transforms.
\Cref{tab:latencywithchannel} summarizes how our method performs in various standards.
Due to space constraints, we only include LIC models with the lowest request latency (Minnen-MSHP) or the lowest compression rate (Chen-BJHAP). Still, with \Cref{tab:infcompexecnobw} and the previous results, we can infer that the LIC baselines have considerably higher latency than FrankenSplit. 

\begin{figure}[htbp]
    \centering
    \includegraphics[width=\columnwidth]{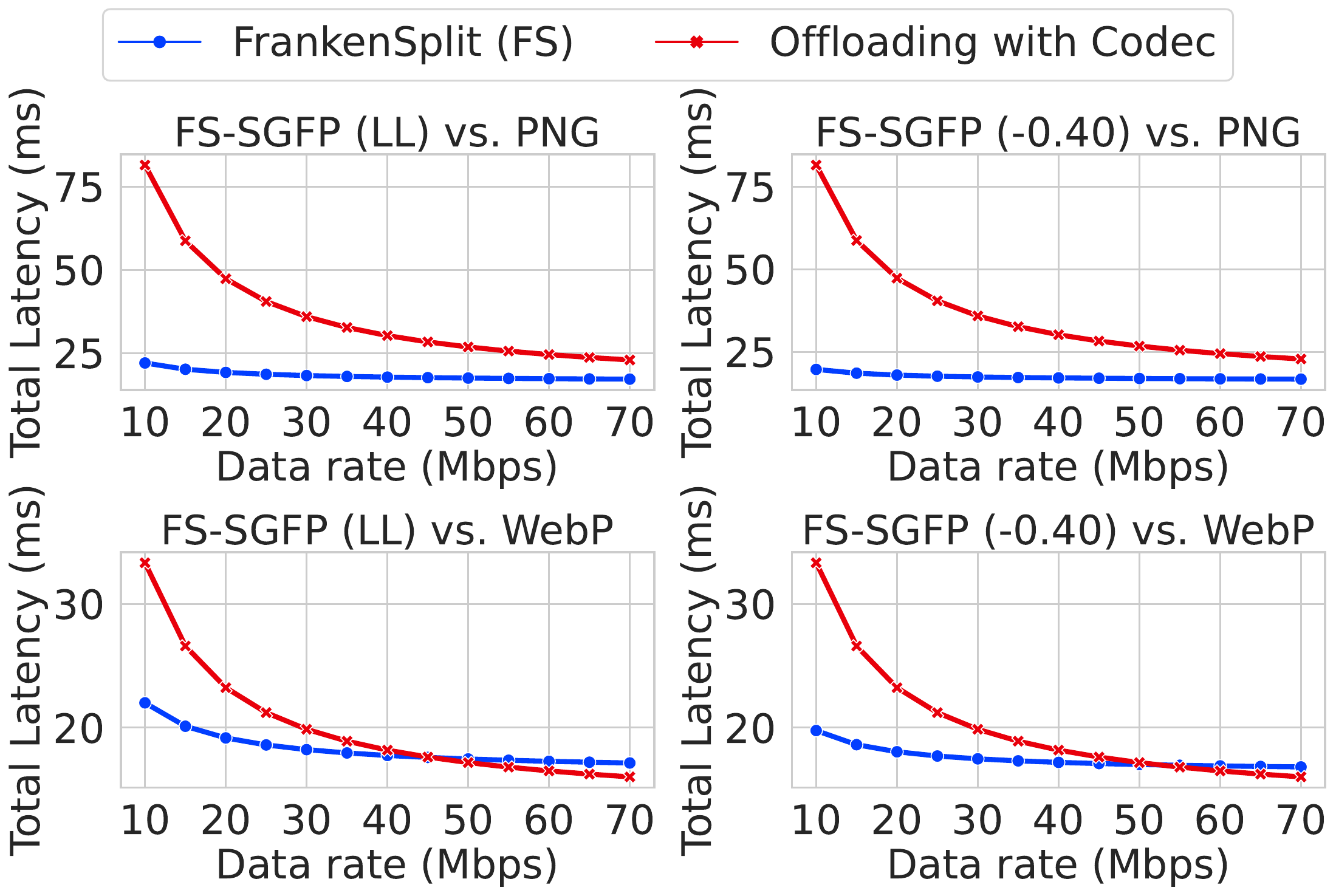}
    \caption{Comparing effects on sizes}
    \label{fig:fsvsoffloading}
\end{figure}

Generally, the more constrained the network is the more we can benefit from reducing the transfer size. In particular, FrankenSplit is up to 16x faster in highly constrained networks, such as BLE. Conversely, offloading with fast handcrafted codecs may be preferable in high-bandwidth environments. Yet, FrankenSplit is significantly better than offloading with PNG, even for 5G. 
\Cref{fig:fsvsoffloading} plots the inference latencies against handcrafted codecs using the NX client. 
For stronger connections, such as  4G LTE, it is still 3.3x faster than using PNG. Nevertheless, compared to WebP, offloading seems more favorable when bandwidth is high. Still, this assumes that the rates do not fluctuate and that the network can seamlessly scale for an arbitrary number of client connections. Moreover, we did not apply any optimizations to the encoder.
\section{Conclusion} \label{sec:conc}
\noindent This work introduced a novel lightweight compression framework to facilitate critical MEC applications relying on large DNNs. We showed that a minimalistic implementation of our design heuristic is sufficient to outperform numerous baselines. However, there are several limitations. We emphasize that the primary insight of the reported results is the potential of adequate distortion measures and regularization methods for neural feature compression. Despite significantly improving rate-distortion performance, better methods may exist to extract saliency maps. 
Moreover, the Factorized Prior entropy model does not discriminate between inputs. Although side information with hypernetworks taken from LIC trivially improves rate-distortion performance, our results show that it may not be a productive approach to repurpose existing image compression methods directly. Hence, conceiving an efficient way to include task-dependent side information is a promising direction.
\bibliographystyle{ieeetr}
\bibliography{main}
\end{document}